\documentclass[twoside,11pt]{article}

\usepackage[accepted]{melba}

\usepackage{amsmath,amsfonts}

\usepackage{hyperref}
\usepackage{url}
\usepackage{subcaption}
\usepackage{caption}
\usepackage{amsmath}
\usepackage{bbm}
\usepackage{multirow}
\usepackage{booktabs}
\usepackage{listings}
\usepackage{tabularx}
\usepackage{booktabs}
\usepackage{float}
\usepackage{multirow}
\usepackage{adjustbox}
\usepackage{xcolor}
\usepackage[normalem]{ulem}
\usepackage{bigdelim}
\usepackage{rotating}

\usepackage{etoolbox}
\usepackage[round-mode=places,detect-weight=true,detect-inline-weight=math]{siunitx}

\usepackage{changes}

\newcolumntype{C}{>{$}c<{$}} 

\newcommand{\newtext}[1]{#1}
\newenvironment{newtextblock}%
  {}%
  {}

  {\color{red}}%
  {}

\melbaheading{2022:023}{https://www.melba-journal.org/papers/2022:023.html}{2022}{1-33}{11/2021}{08/2022}{Stacke, Unger, Lundstr{\"o}m, and Eilertsen}{}{}

\ShortHeadings{Learning Representations}{Stacke, Unger, Lundstr{\"o}m, and Eilertsen}
\firstpageno{1}

\title{Learning Representations with \\ Contrastive Self-Supervised Learning for \\ Histopathology Applications}

\author{\name Karin Stacke \email karin.stacke@liu.se \\  
	\addr Department of Science and Technology (ITN), Linköping University, Sweden \\
	      Sectra AB, Linköping
	\AND
	\name Jonas Unger \email jonas.unger@liu.se \\
	\addr Department of Science and Technology (ITN), Linköping University, Sweden \\
	      Center for Medical Image Science and Visualization (CMIV), Linköping University, Sweden
	\AND
	\name Claes Lundström \email claes.lundstrom@liu.se \\
	\addr Department of Science and Technology (ITN), Linköping University, Sweden \\
	      Center for Medical Image Science and Visualization (CMIV), Linköping University, Sweden \\
	      Sectra AB, Linköping
	\AND
	\name Gabriel Eilertsen \email gabriel.eilertsen@liu.se \\
	\addr Department of Science and Technology (ITN), Linköping University, Sweden \\
	      Center for Medical Image Science and Visualization (CMIV), Linköping University, Sweden
}

\begin{document}

\maketitle

\begin{abstract}
Unsupervised learning has made substantial progress over the last few years, especially by means of contrastive self-supervised learning. The dominating dataset for benchmarking self-supervised learning has been ImageNet, for which recent methods are approaching the performance achieved by fully supervised training. The ImageNet dataset is however largely object-centric, and it is not clear yet what potential those methods have on widely different datasets and tasks that are not object-centric, such as in digital pathology. While self-supervised learning has started to be explored within this area with encouraging results, there is reason to look closer at how this setting differs from natural images and ImageNet. In this paper we make an in-depth analysis of contrastive learning for histopathology, pin-pointing how the contrastive objective will behave differently due to the characteristics of histopathology data. \newtext{Using SimCLR and H\&E stained images as a representative setting for contrastive self-supervised learning in histopathology,} we bring forward a number of considerations, such as view generation for the contrastive objective and hyper-parameter tuning. In a large battery of experiments, we analyze how the downstream performance in tissue classification will be affected by these considerations. The results point to how contrastive learning can reduce the annotation effort within digital pathology, but that the specific dataset characteristics need to be considered. To take full advantage of the contrastive learning objective, different calibrations of view generation and hyper-parameters are required. Our results pave the way for realizing the full potential of self-supervised learning for histopathology applications.
Code and trained models are available at~\url{https://github.com/k-stacke/ssl-pathology}.
\end{abstract}


\section{Introduction}

Deep learning has in the last decade shown great potential for medical image-analysis applications \citep{litjens_2017}. However, the transition from research results to clinically deployed applications is slow. One of the main bottlenecks is the lack of high-quality labeled data needed for training models with high accuracy and robustness, where annotations are cumbersome to acquire and relies on medical expertise~\citep{stadler_2021}.
An active research field has therefore been focused on reducing the dependency of labeled data. This can, for example, be accomplished through \textit{transfer learning}~\citep{yosinski_2014}, where training on a widely different dataset, such as ImageNet~\citep{deng_2009}, can reduce the amount of training data needed in a targeted downstream medical imaging application~\citep{truong_2021}. However, as ImageNet contains natural images, there is a large discrepancy between the source and target domains in terms of colors, intensities, contrasts, image features, class distribution, etc. It has been shown that a closer resemblance between the source and target datasets is preferable~\citep{cui_2018,cole_2021,li_2020d}, and the usefulness of ImageNet for pre-training in medical imaging has been questioned~\citep{raghu_2019}. In addition, ImageNet pre-training has recently also been questioned because of the biased nature of the dataset, intensifying the need to find alternative pre-training methods that are better tailored for the target application~\citep{birhane_2021,yang_2021}.  

Self-supervised learning (SSL) has recently emerged as a viable technique for creating pre-trained models without the need for large, annotated datasets. Instead, pre-training is performed on unlabeled data by means of a proxy objective, for which the labels can be automatically generated. The objective is formulated such that the model learns a general understanding of image content. It can include predicting image rotation~\citep{gidaris_2018}, solving jigsaw puzzles~\citep{noroozi_2017}, re-coloring gray-scale images~\citep{larsson_2017}, to mention a few. 
One family of SSL methods are those based on a contrastive training objective, which is formulated to contract the representation of two positive views, while simultaneously distracting the representations of negative views. This training strategy has shown great promise in the last few years, and include methods such as CPC~\citep{oord_2019}, SimCLR~\citep{chen_2020}, CMC~\citep{tian_2020}, and MoCo~\citep{he_2020}. However, successful results have primarily been presented on ImageNet, despite the above-mentioned need for moving away from this dataset, and it is still unclear how well the results generalize to datasets with different characteristics. 

In this paper, we investigate how SimCLR~\citep{chen_2020} can be extended to learn representations for histopathology applications. \newtext{We consider H\&E stained images, and} take a holistic approach, comparing how differences between the ImageNet dataset and histopathology data influence the SSL objective, as well as pin-pointing how the different components of the objective contribute to the learning outcome. We show that the heuristics that have been demonstrated to work well for natural images do not necessarily apply in histopathology scenarios. \newtext{To explore the differences, we setup} a rigorous experimental study, which includes:
\newtext{
\begin{itemize}
    \item three different pathology datasets, for investigating different scenarios with in-domain and out-of-domain histology data,
    \vspace{-0.1cm}
    \item comparison between ImageNet pre-training and domain-specific SSL, for evaluating the added value of unsupervised pre-training on histology data,
    \vspace{-0.1cm}
    \item evaluation of different data availability scenarios in training of the target application, as we expect pre-training to add different value in the different scenarios,
    \vspace{-0.1cm}
    \item evaluation of the impact of different hyper-parameters, as these have been demonstrated important for the success of contrastive SSL,
    \vspace{-0.1cm}
    \item a study of the convergence behavior of SimCLR on pathology data, for demonstrating how well contrastive pre-training contribute to the target downstream application of tissue classification.
\end{itemize}
}
Our main motive is to clarify how contrastive SSL for histopathology cannot be considered under the same assumptions as for natural image data. Our results lead to a number of important conclusions. For example, we show that:

\begin{itemize}
\item In pathology, SimCLR pre-training gives substantial benefits, if used correctly.
\vspace{-0.1cm}
\item Different types of positive/negative views are optimal for contrastive SSL in histopathology compared to natural images, and the optimal views can be dataset dependent even within the pathology domain.
\vspace{-0.1cm}
\item Parameter tuning, such as the batch size used for SSL, does not have the same influence as for natural images, due to the differences in data characteristics.
\vspace{-0.1cm}
\item Pre-training data aligned with the target pathology sub-domain is better suited compared to more diverse pathology data.
\end{itemize}

\newtext{The paper is organized as follows:
In Section~\ref{sec:related_work}, we discuss and position our work in relation to previous work on SSL, contrastive SSL, and the aspects of SSL specific to digital pathology.
In Section~\ref{sec:bg}, we briefly explain contrastive SSL, followed by a thorough discussion around different aspects related to the view generation that we believe to be important when comparing natural images to digital pathology. This discussion sets the stage for the experimental study performed in Section~\ref{sec:method}, with results presented in Section~\ref{sec:exp}. Finally, in Section~\ref{sec:discussion},
}
we conclude with an outlook on what needs to be considered for further improving contrastive SSL in histopathology, where we emphasize how the differences in data characteristics require a significantly different approach to formulating the contrastive learning objective. We believe that this work is important for broadening the understanding of self-supervised methods, and how the intrinsic properties of the data affect the representations. Our hope is that this will be a stepping stone towards pre-trained models better tailored for histopathology applications.

\section{Related Work}\label{sec:related_work}
A large body of literature has been devoted to unsupervised and self-supervised learning. For self-supervised learning, multiple creative methods have been presented for defining proxy objectives and performing self-labelling. These include, but are not limited to, colorization of grayscale images~\citep{zhang_2016, larsson_2017}, solving of jigsaw puzzles~\citep{noroozi_2017}, and prediction of rotation~\citep{gidaris_2018}. 

Within self-supervised learning, significant attention has recently been given to a specific family of methods, which performs instance discrimination~\citep{dosovitskiy_2014,wu_2018} through contrastive learning with multiple views. 
\cite{bachman_2019} presented a contrastive self-supervised method (AMDIM) based on creating multiple views using augmentation. \cite{oord_2019} presented the InfoMax objective, and showed that by minimizing it you can maximize the mutual information between views. Building on these works, constrastive self-supervised methods such as CMC~\citep{tian_2020}, MoCo(v2)~\citep{he_2020, chen_2020a}, and SimCLR~\citep{chen_2020} have recently shown improved results on ImageNet benchmarks, closing the gap between supervised and unsupervised training. \cite{falcon2020framework} actually showed that many of these methods (such as AMDIM, CPC and SimCLR) are special cases of a general framework for contrastive SSL. As a continuation of this development, methods such as BYOL~\citep{grill_2020} and SwAV~\citep{caron_2021} have been presented, expanding the concept to either avoiding contrastive negatives or to doing cluster assignments instead of instance discrimination.
In this work, we use SimCLR as a representative method of the contrastive self-supervised methodology. \newtext{The choice of method is motivated by its simplicity and high level of adoption since its introduction. We believe that the integral components of contrastive learning can be justly studied through SimCLR, and anticipate that the results will generalize to other state-of-the-art contrastive SSL methods.} 


\begin{newtextblock}
Self-supervised methods have also been applied to medical images in general, and histopathology in specific. 
A number of methods for self-supervised learning have been presented for data such as volumetric CT and MRI, X-ray images and dermatological digital images, with application within classification, localization and segmentation \citep{liu_2019, yan_2020, chaitanya_2020, xie_2020a, zhou_2020a, azizi_2021a, li_2021e, sowrirajan_2021, you_2022}. Many of the works shows that in-domain pre-training is superior to ImageNet pre-training, and that domain-specific selections of positive and negatives views boosts performance. This motivates us to further understand what (if any) considerations that are needed for the domain of histopathology. 

Self-supervised methods developed for histopathology has been presented. For example, incorporation of the spatial information of patches~\citep{gildenblat_2020, li_2021c}, using augmentations based on stain separation~\citep{yang_2021a}, using transformer architectures to capture global information~\citep{wang_2021}, or utilizing the multi-resolution structure of whole-slide images~\citep{koohbanani_2020, srinidhi_2022}. Furthermore, a number of previous work have evaluated contrastive SSL methods that were designed for natural images \citep{lu_2019, stacke_2020, dehaene_2020, ciga_2022}, all showing promising results. Among these, \cite{ciga_2022} are the one closes to this work, as SimCLR is the method used in both studies. Complimentary to \cite{ciga_2022}'s large battery of experiments, we deepen the understanding through experiments regarding optimal view generation and hyper-parameters, in junction with a in-depth discussion on the unique characteristics of histopathology datasets that impact the learned representations.
\end{newtextblock}


Some works give a more rigorous theoretical background to the contrastive methods, such as \cite{arora_2019}, \cite{tsai_2021}, \cite{wu_2020} and \cite{tschannen_2020}. However, much of the success of the previously mentioned methods is derived from heuristics that are still left to be explained theoretically. It is not clear how well the performance showed on one domain transfers to new and different ones. As \cite{torralba_2011} pointed out some time ago, all datasets, ImageNet included, encompasses specific biases that may be inherited by a model trained on the data. \cite{purushwalkam_2020}, for example, argued that the object-centric nature of ImageNet is the reason for why SSL methods based on heavy scale augmentations perform well, but that this approach does not work for object recognition tasks. \cite{cole_2021} showed that contrastive learning is less suited for tasks requiring more fine-grained details, and that pre-training on out-of-domain data gives little benefit. Therefore, we have reason to look closely on how contrastive learning methods transfer to the domain of histopathology.

\section{Background}
\label{sec:bg}
This section gives an overview of contrastive multi-view learning as well as view generation, with the goal of giving an conceptual description of how different design choices affect the learned representation. \newtext{The descriptions will facilitate the experimental design and result analysis in Section~\ref{sec:method}-\ref{sec:discussion}, for identifying the differences in contrastive SSL for histopathology compared to object-centered datasets with natural images.}

\subsection{Contrastive learning}
\label{sec:theory}
The general idea of contrastive learning is that an \textit{anchor} data point $x_i$ (sometimes referred to as query) together with a \textit{positive} data point $x_j$ (key) form a set of positive \textit{views} of the same object. The goal is to map these views to a shared representation space, such that the representations contain underlying information (features) shared between them, while simultaneously discarding other (nuisance) information. A positive view pair could therefore share the information of depicting the same object, but may differ in view angle, lighting conditions, or occlusion level. 

For images, this shared information is high-dimensional, which makes its estimation challenging. The views are therefore encoded to a more compact representation using a non-linear encoder, $g$, $\{z_i = g(x_i), z_j = g(x_j)\}$, such that the mutual information between $z_i$ and $z_j$ is maximized. Maximization of the mutual information can be estimated by using a contrastive loss function~\citep{oord_2019}, defined to be minimized by assigning high values of positive pairs of data ($\{z_i, z_j\}$) and low values to all other ($\{z_i, z_k\}, k \neq \{j,i\}$ (denoted “negatives”). A popular such loss function is the InfoNCE loss, defined as:
\begin{equation}
\mathcal{L}_i = - \log \frac{\exp{(z_i^T z_j)}}{\sum_{k} \exp{(z_i^T  z_k)}}.
\end{equation}
The choice of positives can be done either in a supervised way, where coupled data is collected (such as multiple staining of the same tissue sample), or in a self-supervised manner, where the views are automatically generated. One popular approach of the latter kind is to create two views from the same data point, $x$, by applying random transformations, $\{t_i \sim T, t_j \sim T, \}$, such that two views of the data sample are created, $\{\widetilde{x}_i, \widetilde{x}_j\}$. Negative samples are typically taken as randomly selected samples from the training data. \textit{View generation}, that is, how the views are chosen, has a direct impact of what features the model will learn.

\subsection{View Generation}
\label{sec:viewgen}

Due to the way the contrastive objective is formulated, the choices of how positive and negative views are selected will largely impact the learned representation. If they are chosen correctly, the learned representation will separate data in such a way that is useful for the target \textit{downstream} task. Incorrectly chosen, the model may learn \textit{nuisance} features that result in poor separation of the data with respect to the downstream task. The choice of optimally selecting positives and negatives thus depends on the intended downstream task, since it needs to take into account what is considered \textit{task relevant} and \textit{task irrelevant~\footnote{the notation "task irrelevant" and "nuisance" features will be used interchangeably}}~\citep{tian_2020b}. For example, color may be considered a nuisance variable in the downstream task of tumor classification in H\&E slides and should therefore not be shared between positives, but may be an important feature if the downstream task is scoring of immunohistochemical staining. As the SSL is task-agnostic, that is, the downstream task is unknown with regard to the self-supervised objective, the view generation is critical for controlling what features the model learns, such that they tailored to the downstream task.

\begin{figure}[t]
    \centering
    \includegraphics[width=\linewidth]{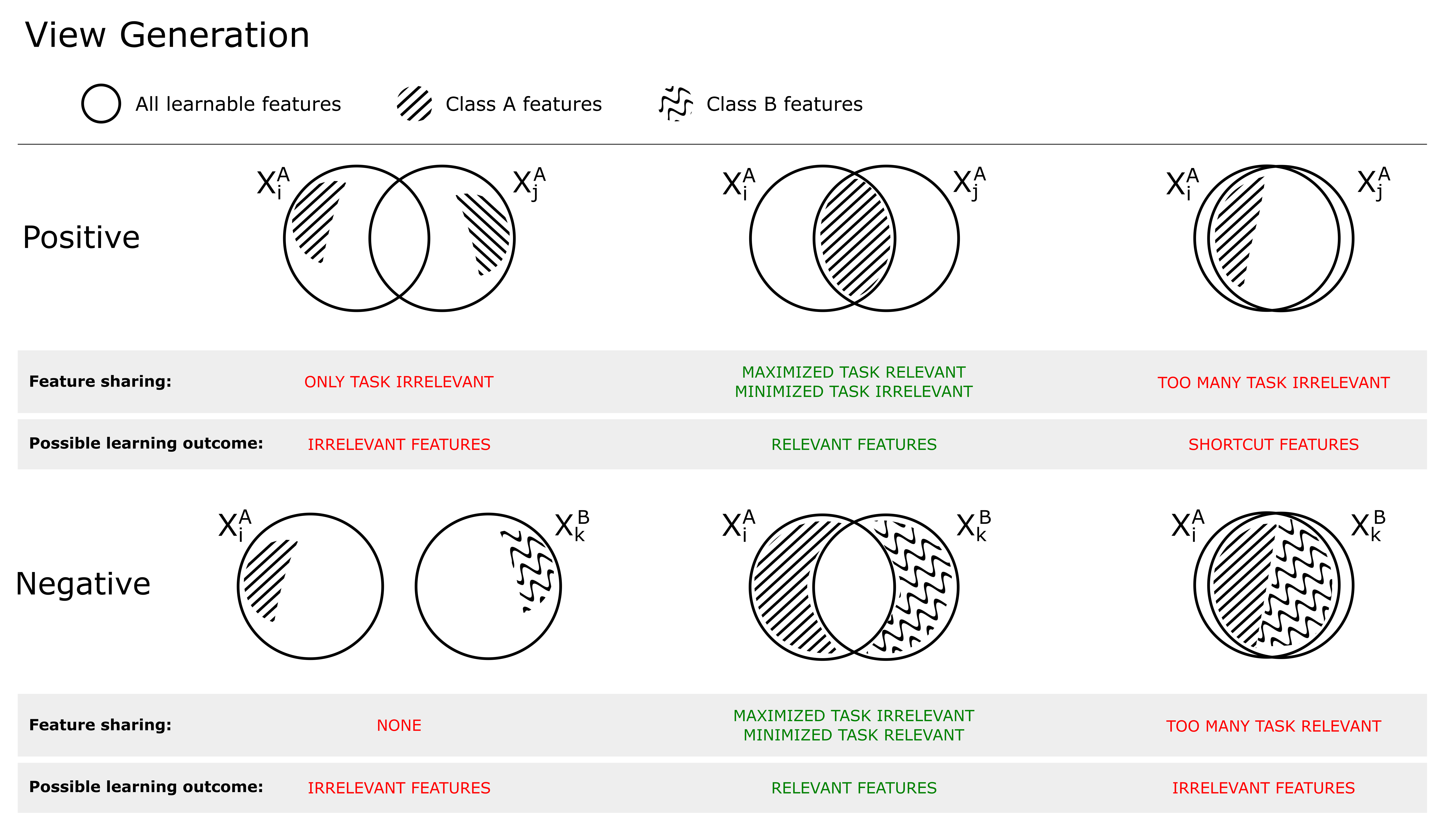}
    \caption{How view generation is done will affect the learning outcome. Circles represent all learnable features for one data sample, where shaded areas denote task-relevant features. Between anchor ($x^A_i$) and positive ($x^A_j$), which belong to the same class (A), sharing of task-relevant features should be maximized, while simultaneously minimizing task-irrelevant features. Between the anchor ($x^A_i$) and the negative ($x^B_k$) (not belonging to same class, $A \neq B$), the opposite should be true.}
    \label{fig:view_gen}
\end{figure}

Figure~\ref{fig:view_gen} shows the relationship between shared mutual information (SSL objective) and view generation. Positive views (top) can be selected/created such that 1) no task-relevant features are shared, resulting in that the model will use only task-irrelevant features to solve the self-supervised objective, 2) only task-relevant features are shared, resulting in the model learning (a subset) of these, or 3) both task-relevant and irrelevant features are shared, increasing the risk of the model learning so called shortcut features \citep{geirhos_2020}, often low-level features (such as color). If the views are created with augmentations (which is what we will consider in this work), this is the result of 1) too strong, 2) just right or 3) too little transformations. To achieve optimal performance on the downstream task, the model should learn the minimally sufficient solution~\citep{tsai_2021}, such that two positive views share as much task-relevant information as possible, and as little task-irrelevant information (middle column, Figure~\ref{fig:view_gen}). 

As highlighted by \cite{arora_2019}, the choice of negatives (which generally are randomly selected from the mini-batch) is also important for the learning outcome. In the bottom row of Figure~\ref{fig:view_gen}, the relationship between an anchor and negative is shown. If no information is shared, the model does not have to learn any task-relevant features as any feature may solve the pre-training objective (left). If too much information is shared, the model will not learn task-relevant features as these cannot be used to distinguish between positive and negatives (right). This is typically the case when negatives belong to the same (latent) class as the positive, making them so called \textit{false negatives}. \newtext{It is important to distinguish between \textit{false} negatives and \textit{hard} negatives, where hard negatives are true negatives which share similar features with the anchor. This is shown in the middle column, where the shared information between target and negative is composed of substantial amount of nuisance information, but no task related information. Hard negatives are generally beneficial for the learning outcome~\citep{robinson_2021}, as they hinder the model to rely on task-irrelevant features to solve the contrastive objective.}

\begin{figure}
    \centering
    \includegraphics[width=\textwidth]{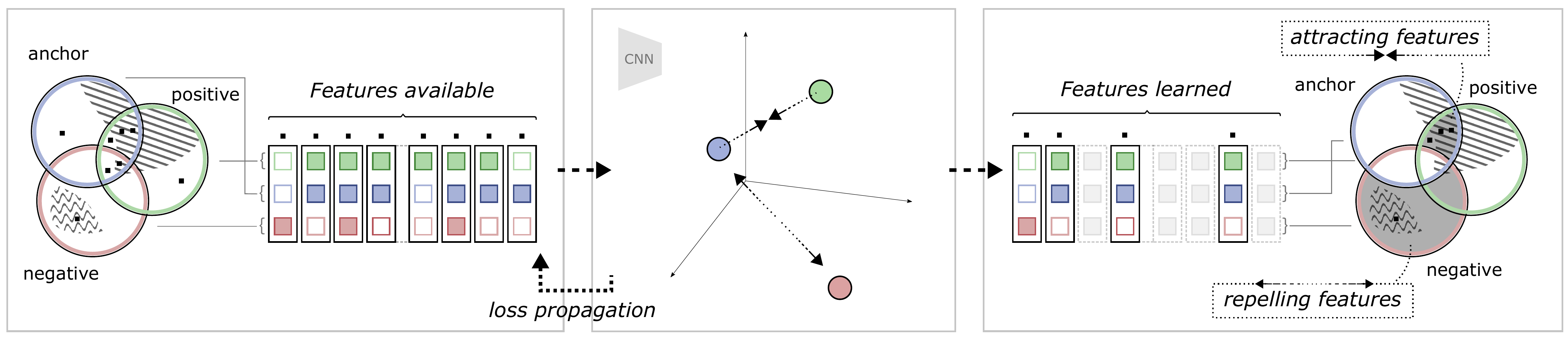}
    \caption{View generation results in shared features between anchor, positive and negative views. Due to the contrastive objective, only a subset of all available features are learned by the model. These features are either \textit{attracting} features or \textit{repelling} features, where view generation is the only means of controlling to what extent these features are task-relevant.}
    \label{fig:contrastive_objective}
\end{figure}

To further illustrate the relationship between the different views and the self-supervised objective, an example is shown in Figure~\ref{fig:contrastive_objective}. In this example, view generation resulted in some features shared between the anchor and positive views, of which a subset are task-relevant. Some task-relevant features are, however, not shared, existing in only one of the views. In addition, some features (both task-relevant and irrelevant) are also shared by a negative view. This means that out of all available features, some exist in one, two or all of the views. During model training, the model learns to represent each data point such that the contrastive objective is fulfilled: the anchor-positive views are attracted, and the anchor-negative views are repelled. The \textit{attracting features} are found in the intersection between the anchor and positive but not in the negative. The \textit{repelling features} are found in the negative, but not in the anchor. As discussed in the previous section, the region of attracting features should therefore contain primarily task-relevant features, and the repelling region should only contain task-irrelevant features. It should, however, be noted that there is no guarantee that the model will learn all features in these regions, but only the subset of features that is enough to solve the contrastive objective, as observed by~\cite{tian_2020b}. In the end, as the contrastive objective is task-agnostic and completely relies on this distinction of features over the training dataset, the degree of task relevance of the learned representation will depend on how the views were generated.

\section{Method}
\label{sec:method}
The goal of this paper is to better understand how contrastive self-supervised learning (SSL) can be used for clinical applications where labeled data is scarce. We do this by evaluating different pre-training methods for the target downstream application, i.e., classification with varying amounts of labeled data. In doing so, it is necessary to systematically analyze and understand the impact of different pre-training methods and training strategies, and how this relates to the type of data used and the target task. In this section, we present the datasets, training details and evaluation metrics.  

\subsection{Experiment design}
\label{sec:exp_design}
By using SimCLR as a representative method for contrastive learning, we build our investigation on a series of experiments where we vary the parameters and methodologies relevant to the analysis. 

\begin{itemize}
    \item \emph{Training SimCLR models on in-domain and out-of-domain histology data}. It is of key interest to understand how pre-training from different histopathology sources and with different augmentations, affect the resulting learned outcome. The results from this analysis will form the basis of our discussion on self-supervised learning for histopathology data. 
   
    \item \emph{Compare domain-specific SimCLR with ImageNet pre-training and no pre-training}. Previous works often rely on transfer learning from pre-training using ImageNet. Motivated by the strong differences between pathology data and ImageNet data in terms of e.g., image content, number of classes, and overall composition, a systematic evaluation is done of whether a domain-specific pre-training using SimCLR is more beneficial in this context and if so why.

    \item \emph{Evaluation of different amounts of supervised data for downstream-task training}. Pre-trained models are evaluated both with respect to linear and fine-tuning performance with varying amounts of supervised training data. 
    
    \item \emph{Batch size, learning rate \newtext{and temperature scaling} impact}. Tuning of hyper-parameters such as batch size, learning rate\newtext{, and temperature scaling} have been shown to play an important role in contrastive learning using ImageNet. This experiment explores the corresponding parameter tuning for histopathology data.  
    
    \item \emph{Evaluation of performance during training}. Training dynamics presents important information on model robustness, the optimization, and overall performance. This experiment investigates downstream task training and how the resulting models evolve over time.

\end{itemize}

These experiments, conducted with multiple datasets, form the basis for a detailed evaluation of contrastive self-supervised learning in general, and SimCLR in particular, in the context of histopathology.

\subsection{Datasets}
\label{sec:datasets}

\begin{figure}
    \centering
    \begin{subfigure}{.22\textwidth}
        \centering
        \includegraphics[width=\linewidth]{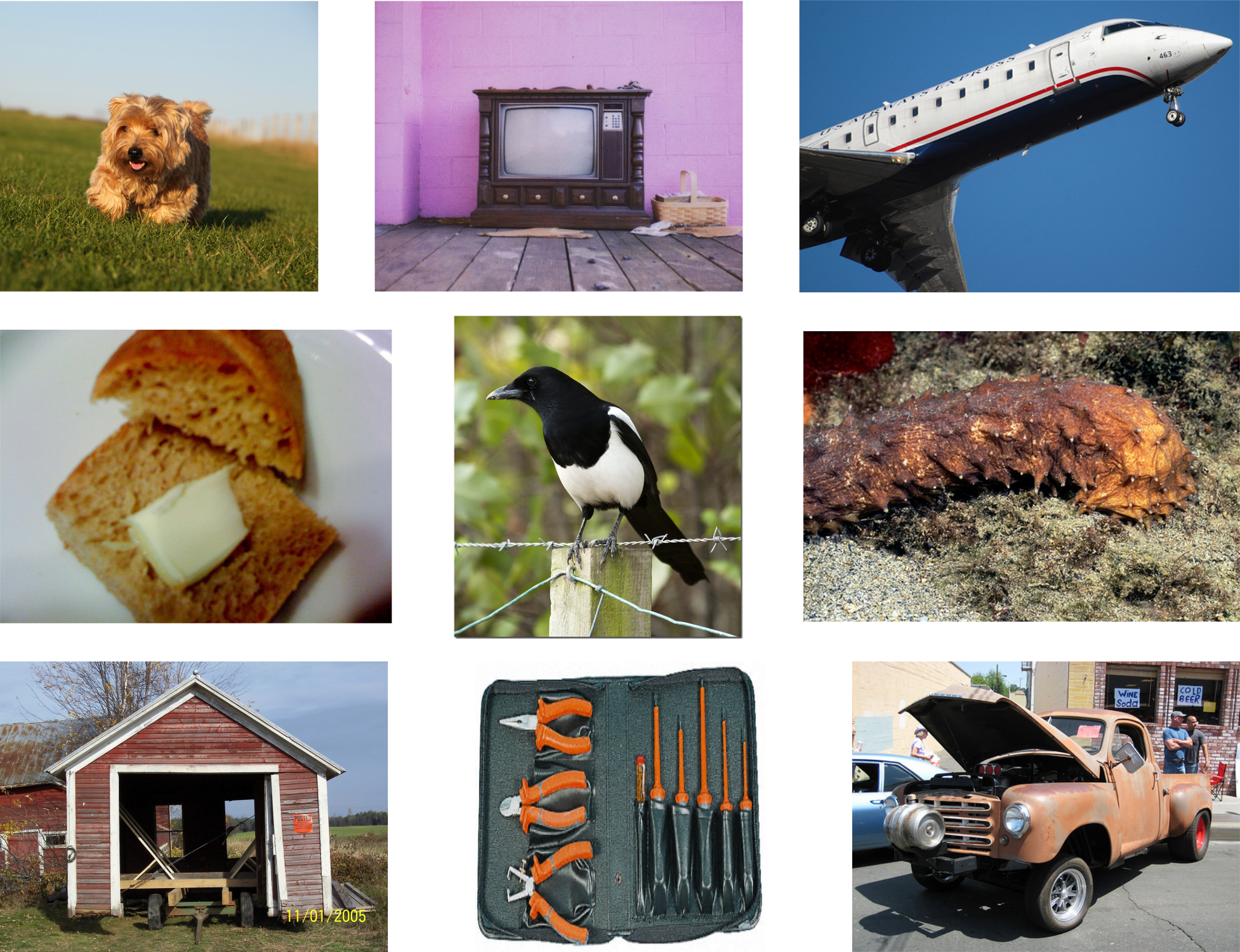}
        \caption{ImageNet}
    \end{subfigure}%
    \hfill
    \begin{subfigure}{.17\textwidth}
        \centering
        \includegraphics[width=0.95\linewidth]{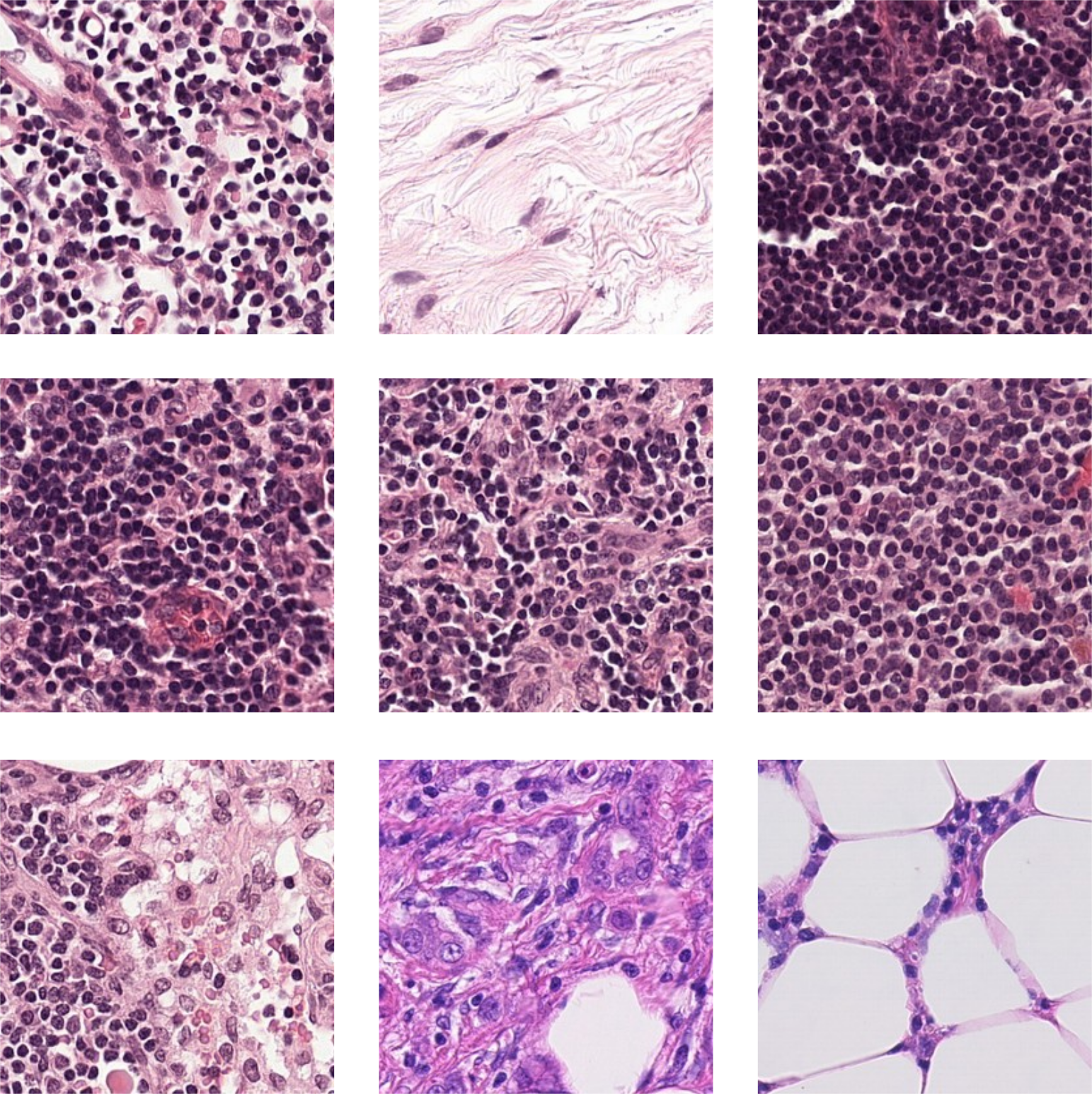}
        \caption{Camelyon16}
    \end{subfigure}%
    \hfill
    \begin{subfigure}{.17\textwidth}
        \centering
        \includegraphics[width=0.95\linewidth]{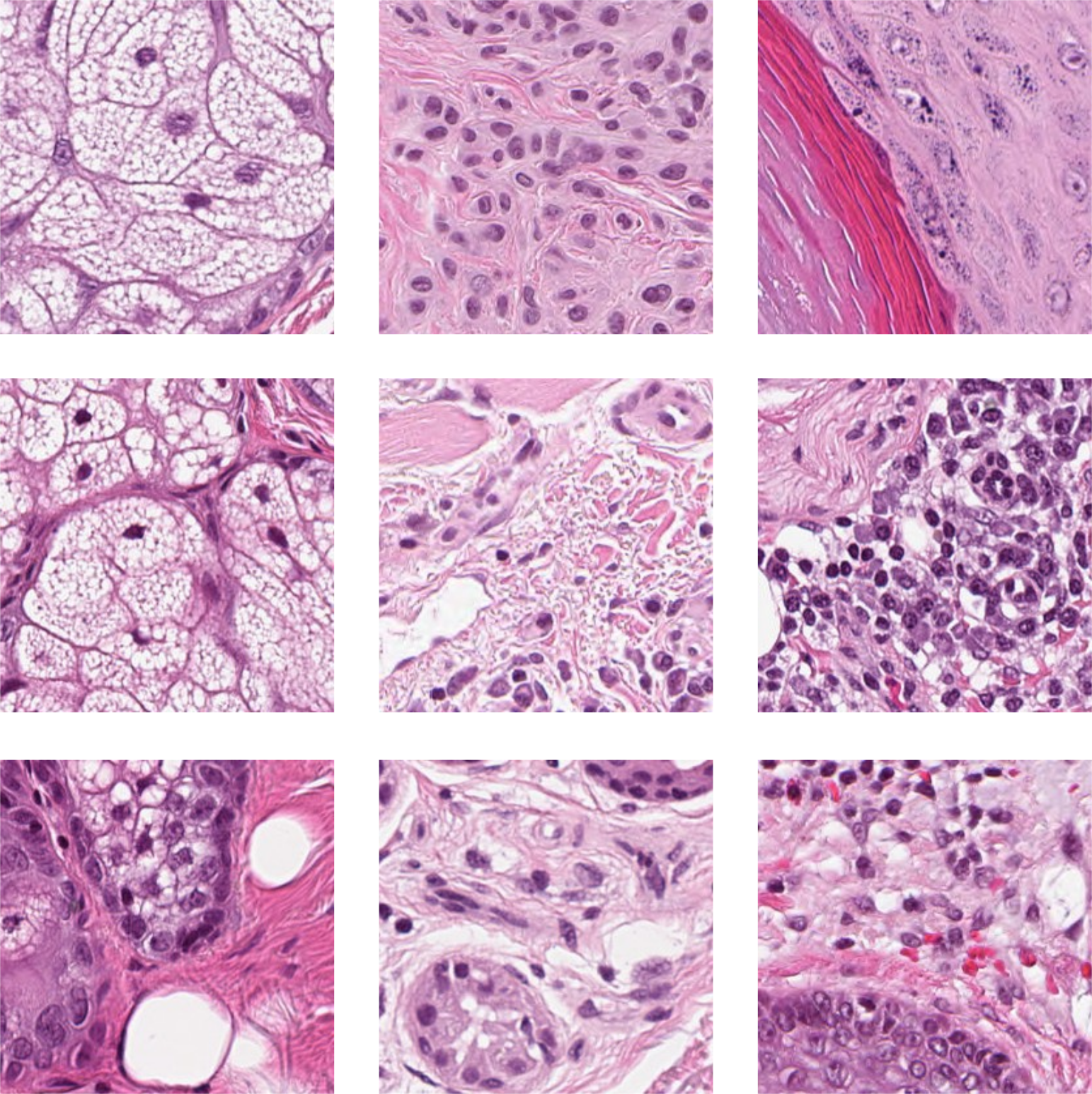}
        \caption{AIDA-LNSK}
    \end{subfigure}%
    \hfill
    \begin{subfigure}{.17\textwidth}
        \centering
        \includegraphics[width=0.95\linewidth]{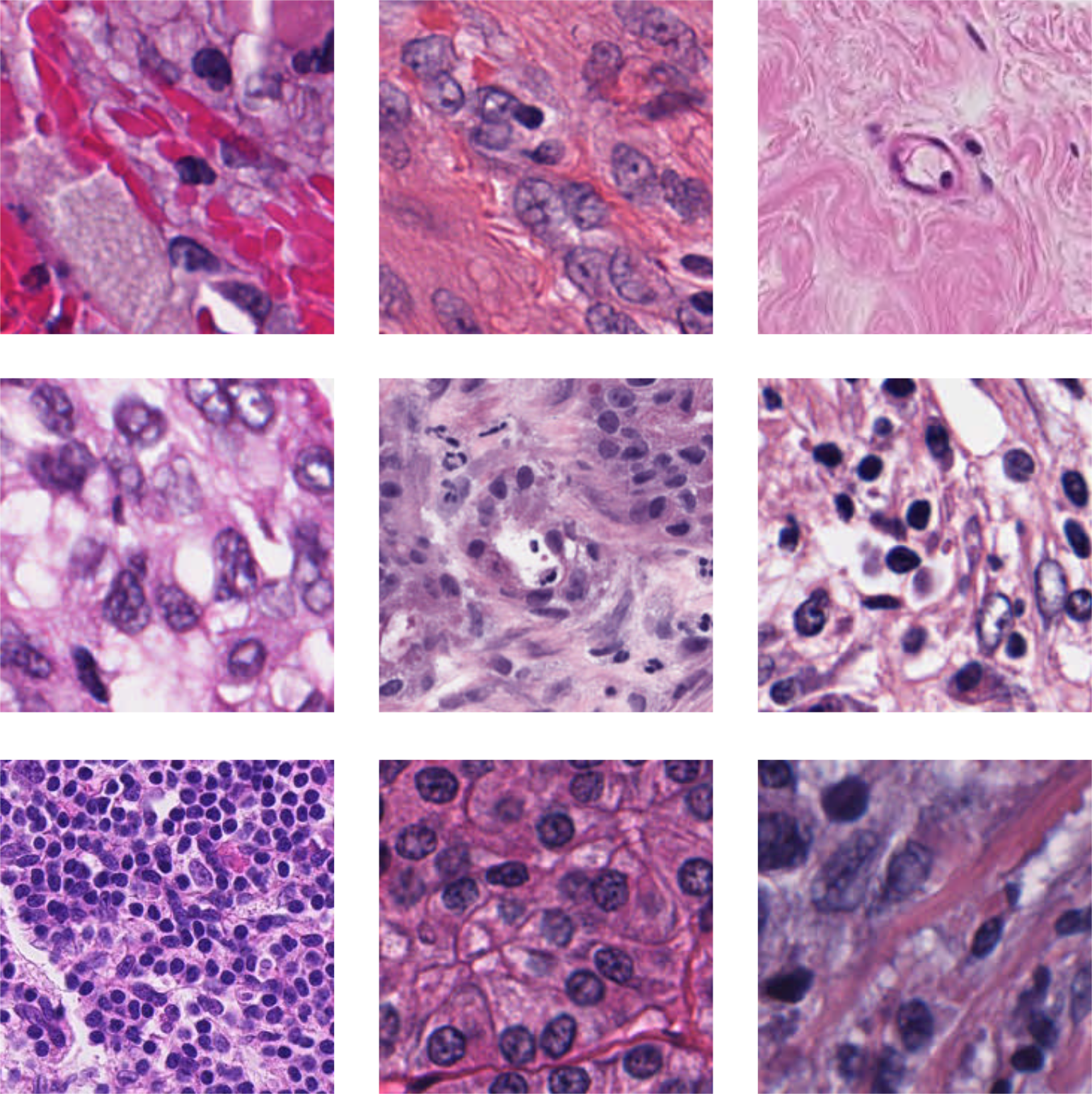}
        \caption{Multidata}
    \end{subfigure}%
    \caption{Example images from datasets.}
    \label{fig:example_images}
\end{figure}

For this study, three different histopathology datasets were used, from sentinel breast lymph node tissue, skin tissue and one consisting of mixed tissues extracted as a subset from 60 different datasets. As reference dataset, ImageNet is used. Examples images are shown in Figure~\ref{fig:example_images}, with further details given below, and in Appendix~\ref{sec:appendix_data}. 

\paragraph{ImageNet ILSVRC2012}~\citep{deng_2009}: a dataset constructed by searching on the internet using keywords listed in the WordNet database. A subset of the total dataset, approximately 1.2M images, was labeled as belonging to one of 1000 classes using a crowd sourcing technique. Despite the aim of being a representative dataset with a wide category of objects, the nature of the collection technique and annotation strategy has resulted in distinct characteristics and biases in the data, resulting in models trained on this dataset may inherit the biases~\citep{torralba_2011}. SSL methods developed and tested on this dataset are therefore also likely to adhere to some inherit characteristics of the data~\citep{cole_2021}. Using ImageNet pre-trained weights for transfer learning is a common approach for many medical image applications, which motivates us to use it as a baseline method. Pre-trained models (trained supervised) were accessed though the Pytorch library~\footnote{Accessible here: \url{https://pytorch.org/vision/stable/models.html}}. 

\paragraph{Camelyon16}~\citep{litjens_2018}:  399 H\&E-stained whole-slide images (WSIs) of sentinel lymph node tissue, annotated for breast cancer metastases. This dataset was sampled into smaller patches twice, to construct one dataset used for self-supervised training, and one for supervised training. 
For unsupervised training, the WSIs were sampled in an unsupervised way, i.e., no tissue annotations were used to guide sampling from the 270 slides selected as the training slides from the official split. Patches were sampled non-overlapping with patch size of 256x256 pixels with a resolution of 0.5 microns per pixel (mpp) (approximately 20x). Maximum 1000 samples were chosen per slide, resulting in a dataset consisted of slightly less than 270k images.

For supervised training, a downstream task was formulated as binary tumor classification task. For this dataset, which we denote \textit{PatchCamelyon20x}, the patches were sampled in accordance to the PatchCamelyon~\citep{veeling_2018} dataset, a pre-defined probabilistic sampling of the Camleyon16 dataset using the pixel annotations, resulting in a class-balance between tumor and non-tumor labels in the dataset. The original PatchCamelyon dataset is sampled at 10x, with patch size of 96x96 pixels. In this study \newtext{however}, the data was resampled to match the unsupervised dataset \newtext{at the target resolution of 0.5 microns per pixel (approx. 20x) with patch size 256x256 pixels} (at the same coordinates as the original dataset). In line with the PatchCamelyon dataset, training/validation/test samples for \textit{PatchCamelyon20x} were taken from 216/54/129 slides respectively. In addition, subsets (possibly overlapping) of the supervised training dataset was selected as taking all patches from 10, 20, 50, 100 random slides, respectively (the smaller subsets are subsets of the larger ones). This was repeated five times to create five folds for each subset. For more details about PatchCamelyon, see~\cite{veeling_2018}. Pre-training SimCLR models using the unsupervised dataset is therefore considered \textit{in-domain} pre-training, as the same slides are re-sampled and used for training the supervised, downstream task. 

\paragraph{AIDA-LNSK} \citep{lindman_2019a}: 
a dataset containing 96 WSIs from 71 unique patients of skin tissue. The data was split into train, validation, and test on patient level, such that 50, 6, and 15 patients were included in train, validation and test respectively. This resulted in 65, 8, and 23 WSIs in each dataset. In analogy with Camelyon16, the AIDA-LNSK is sampled to create two dataset, one for downstream task training, and a corresponding \textit{in-domain} dataset for pre-training. 

For unsupervised training, patches were extracted from tissue regions of slides in the training set (65 slides), found by Otsu threshold from WSI magnification 5x. From these regions, the data was sampled without overlap. This resulted in an unsupervised dataset size of approximately 270k patches, roughly the same size as the unsupervised Camelyon16 dataset. All patches were extracted with 0.5 (mpp) resolution at a size of 256x256 pixels.  

From AIDA-LNSK, a downstream task was constructed as a five-class tissue classification task, using available pixel-level annotations. The five classes were formed as four classes representing healthy tissue types (dermis, epidermis, subcutaneous tissue and skin appendage structure) and one class representing “abnormal” (containing different types of cancer, inflammation, scaring and so on). The slides from the above mentioned split was sampled (same size and resolution as the unsupervised dataset) such that for the supervised training dataset, each class included \textit{at least} 75'000 samples, resulting in approximately 320k patches. The training set was thereafter subdivided, by randomly selecting all patches from 10, 20 and 50 slides from the original 65. Smaller subsets are true subsets of larger ones. This was repeated 5 times, such that for each subset size, 5 (possibly overlapping) dataset were created. The validation and test set were sampled from the respective slides in a class-balanced way. For more information about the data collection and annotations, please see \cite{stadler_2021}.

\paragraph{Multidata}: \cite{ciga_2022} constructed a multi-data dataset, consisting of samples from 60 publicly available datasets, originating from multiple tissue types. This pre-sampled dataset was sampled unsupervised, and will in this study be used for self-supervised training only. Patches were extracted with size 224x224 pixels, at the maximum available resolution per dataset, resulting in a variation of resolution between the patches (0.25--0.5 mpp). In this study, we use a 1\% subset of this data, provided by the authors, consisting of 40k patches. With relation to the downstream tasks of breast tumor classification and skin tissue classification, this data is considered \textit{out-of-domain}.

\subsection{Training}
\label{sec:aug}
For all experiments, the ResNet50~\citep{he_2016} model architecture was used. As self-supervised method, SimCLR was evaluated, and if nothing else is stated, the same training setup was used as in \cite{chen_2020}. 

The SimCLR objective is to minimize the NT-Xent loss. For a positive pair this is defined as
\begin{equation}
    \mathcal{L}_{i,j} = -\log \frac{\exp({\text{sim}(z_i, z_j)/\tau)}}{\sum_{k=1}^{2N}\mathbbm{1}_{[k \ne i]} \exp({\text{sim}(z_i, z_k)/\tau) }},
    \label{eq:ntxent}
\end{equation}
where $(i,j)$ is a positive pair, and $(i, k)$ a negative one, with the similarity function defined as $\text{sim}(u,v) = u^T v/(\|u\| \|v\|)$ (cosine similarity). The temperature scaling, $\tau$, was set to 0.5 for all experiments\newtext{, unless otherwise stated}.

In SimCLR, augmentations are used to create the positive views from the same anchor sample. Henceforth, "original" augmentations will refer to the augmentations defined in the SimCLR paper (randomly applying resize and crop, horizontal flip, color jittering and Gaussian blur) with the modification of when training with histopathology data, additional vertical flip and random rotation of 90 degrees was added (due to the rotation invariance of histopathology data). For examples, please see Appendix~\ref{sec:appendix_aug}.

Following commonly used protocol, the self-supervised pre-training was done once (due to computational and time constraints). The resulting representation is evaluated primarly using a linear classifier on top of the frozen weights, but also using fine-tuning (linear classifier on top, without freezing any weights). The former method is a way of evaluating the quality of the pre-trained representations with regard to the target data and objective, while the latter is a more realistic use-case of the trained weights. For supervised training cases, training was repeated 5 times with different seeds, and when subsets of the training data is used, also different folds. All results are reported as patch-wise accuracy on class-balanced test sets. 

All training was conducted on 4 NVIDIA V100 or NVIDIA A100 GPUs. For SimCLR training, effective batch size of 1024 was used for 200 epochs (training time approximately 24 hours). LARS~\citep{you_2017} was used as optimizer, with an initial learning rate $1.2$ regulated with the a cosine annealing scheduler.

For linear evaluation, models were trained in a supervised manner for 20 epochs using Adam optimizer with an initial learning rate of 0.01. For fine-tuning, models were trained for 50 epochs. For breast, Adam optimizer with an initial learning rate of $1\mathrm{e}{-3}$ was used, with weight decay of $1\mathrm{e}{-4}$. For skin, SGD optimizer with Nesterov momentum was used with initial learning rate of $1\mathrm{e}{-4}$ and momentum parameter of 0.9. In addition, models were trained in a supervised manner with random initialization (``from scratch''), using Adam optimizer with learning rate 0.001 for 50 epochs. 
Common for all supervised training was the usage of cosine annealing scheduler to reduce learning rate and weighted sampling to mitigate effects of class imbalance. Augmentations applied during supervised training consisted of: random resize crop with scale variance between 0.95--1.0, color jittering, and rotation/flip.

\section{Results}
\label{sec:exp}
Below follows detailed description and results from the experiments outlined in Section~\ref{sec:exp_design}.

\subsection{Positive-view generation by augmentation}
\label{sec:better_aug}

\begin{table}[b]
\small
\caption{Relative improvement (percentage points) over using base augmentations only (rotate, flip, color jitter). SimCLR model trained for 50 epochs, linear evaluation on supervised training set from 50 slides.}
\centering
\begin{tabular}{llll}
\hline
\textbf{Augmentations}                  & \textbf{Breast} & \textbf{Skin} \\ \hline
\textit{Base}                                    & $81.75 \%$       & $68.3 \%$     \\
\textit{+ \{Gaussian blur\}}                     & + 1.26          &  + 0.37    \\
\textit{+ \{Scale\}}                             & \textbf{+ 3.94} &  + 0.64 \\
\textit{+ \{Gaussian blur, Scale\} (SimCLR Orig.)} & + 0.2         & + 1.25 \\
\textit{+ \{Scale, Grid Distort\}}               & + 2.44          & + 1.42 \\
\textit{+ \{Scale, Grid Distort, Shuffle\}} & + 1.72          & + 1.39 \\
\textit{+ \{Grid Shuffle\}}                      & + 0.37          &  + 2.83   \\
\textit{+ \{Grid Distort, Shuffle\}}        & + 1.6           &  + \textbf{3.79}  \\ \hline
\end{tabular}
\label{tab:aug}
\end{table}

We here investigate the effect of different augmentations, and their ability to isolate the task-relevant features in order to improve the correlation between the SSL objective and the downstream performance. 

SimCLR models were trained on the unsupervised datasets from either breast or skin, and evaluated after 50 epochs on in-domain data from 50 slides, respectively. Eight different augmentation combinations were evaluated in terms of the relative improvement over \textit{Base} augmentations (using flip, rotate, color jitter and very low scale variance, $0.95-1.0$). As \cite{chen_2020} found large scale variance (0.2--1.0) together with Gaussian blur beneficial for ImageNet, these augmentations were evaluated both together and individually. Furthermore, two additional augmentations were evaluated, \textit{Grid Distort} and \textit{Shuffle}. These were chosen as transformations that preserve the label of histopathology patches, but adds perturbations of the compositions of the cells. The results are shown in Table~\ref{tab:aug}. Further details and examples of the augmentations can be found in Appendix~\ref{sec:appendix_aug}. 

\paragraph{Choosing optimal augmentations for histopathology data depend on dataset and downstream task}
Looking at Table~\ref{tab:aug}, choosing the appropriate augmentations for view generation makes it possible to boost performance with $3.94$ and $3.79$ percentage points for breast and skin respectably. However, it appears as there is no common set of augmentations which is optimal for both datasets. Furthermore, using the same augmentations that were presented in \cite{chen_2020} as optimal for ImageNet gives sub-optimal performance for histology data. Using Gaussian Blur was found to be of negligible value, and scale was only substantially beneficial for breast data, not for skin. The best set of augmentations for breast data was to use \textit{Base + Scale}, while for skin, \textit{Base + Grid Distort + Shuffle} gave the highest performance.  

Thus, different sets of augmentations are optimal for different datasets and different downstream task. This is not surprising, as the relevant information in the data depends both on the inherited features in the dataset, as well as what the downstream task is. Finding task- and data-specific augmentations are therefore needed.

\begin{figure}[!b]
    \centering
    \includegraphics[width=\linewidth]{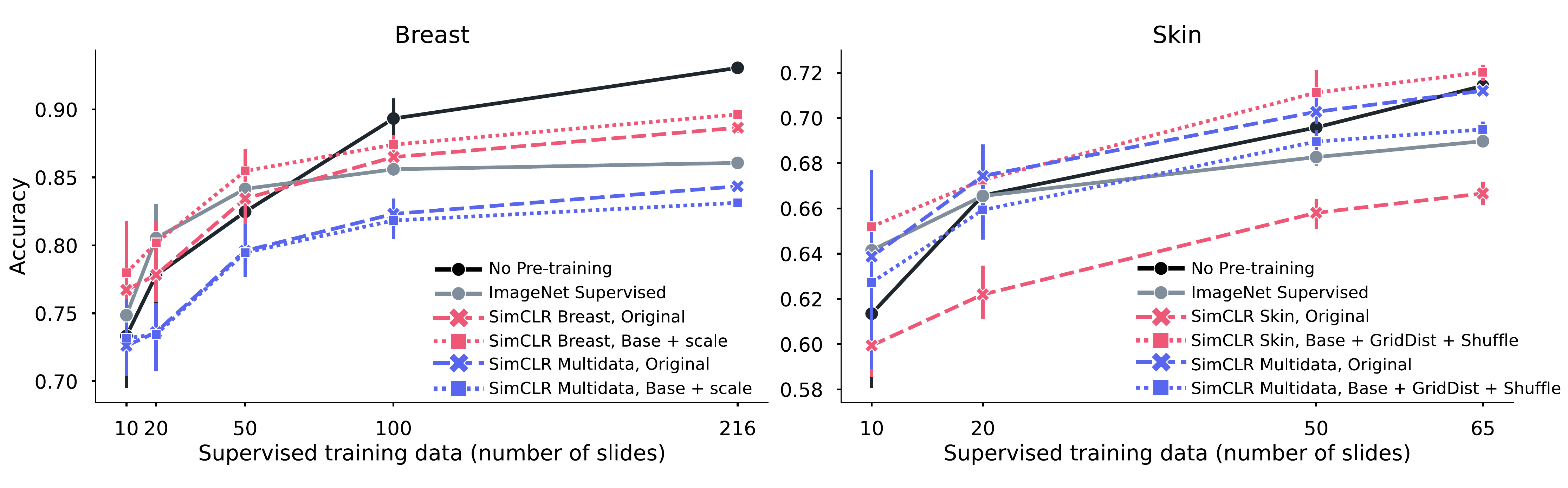}
    \caption{Patch-level performance of linear trainings of different pre-training strategies, at varying supervised training size (number of slides). The linear models are trained on representations learned by: ImageNet supervised pre-training (gray, solid), or using SimCLR pre-training with different augmentations (dashed/dotted), and on different datasets (in-domain pathology data: red, out-of-domain pathology data: blue). The reference (black, solid) is the full ResNet50 model trained in a supervised way on the (subset) training data. Left: breast data, right: skin data. Note different x-axes between subplots.}
    \label{fig:performance}
\end{figure}

\subsection{Downstream Performance}
In this section, different pre-training strategies are evaluated. By using a random initialized model trained in a supervised way as reference, we want to understand the gain of using a pre-trained model depending on the size of the supervised data. In Figure~\ref{fig:performance}, the result of linear evaluation (frozen pre-trained weights) for breast and skin is shown, where the supervised reference is shown in black (solid line). For each tissue type, five different pre-trained models are evaluated, either ImageNet Supervised (gray, solid) or four different configurations of SimCLR, where color denotes dataset (in-domain in red, out-of-domain in blue) and markers denote augmentations applied (original SimCLR in dashed, best set from Table~\ref{sec:better_aug} as dotted). 
Table~\ref{tab:finetuning} shows the fine-tuning results comparing the pre-training method giving the best performance on the linear evaluation with ImageNet pre-training and no pre-training (random initialization). From the results in the figure and table, we can draw a number of conclusions.

\begin{table}[b]
\centering
\caption{Fine-tuning performance (patch-level accuracy, \%) using no pre-training (random initialized weights), ImageNet pre-training or SimCLR in-domain training. 
}
\label{tab:finetuning}
\begin{adjustbox}{width=\textwidth}
\begin{tabular}{@{}llCCCCCC@{}}
\toprule
\textbf{Data} & \textbf{Pre-training} & \multicolumn{6}{c}{\textbf{Supervised size (\#~slides)}} \\ \cmidrule{3-8}
 &  & 10 & 20 & 50 & 65 & 100 & 216 \\ \midrule
\multirow{3}{*}{\textit{Breast}}
 & \textit{None}                        & 73.33 \pm 4.82 & 77.81 \pm 2.27 & 82.47 \pm 0.95 & - & 89.33 \pm 1.73 & 93.06 \pm 0.21 \\
 & \textit{ImageNet Supervised}         & 71.62 \pm 5.74 & 82.67 \pm 3.10 & 89.11 \pm 2.43 & - & 91.35 \pm 0.98 & 93.13 \pm 0.42 \\
 & \textit{\begin{tabular}[c]{@{}l@{}}SimCLR Breast, Base \\ + Scale\end{tabular}} 
                                        & 76.62 \pm 6.92 & 84.58 \pm 2.76 & 90.60 \pm 1.81 & - & 91.63 \pm 1.14 & 92.59 \pm 0.36 \\ \midrule
\multirow{3}{*}{\textit{Skin}}                      
 & \textit{None}                        & 61.35 \pm 3.88 & 66.57 \pm 1.77 & 69.58 \pm 0.66 & 71.42 \pm 0.29 & - & - \\
 & \textit{ImageNet Supervised}         & 64.85 \pm 4.22 & 67.64 \pm 1.27 & 70.23 \pm 0.14 & 70.96 \pm 0.22 & - & - \\
 & \textit{\begin{tabular}[c]{@{}l@{}}SimCLR Skin, Base \\ + Distort + Shuffle\end{tabular}} 
                                        & 66.27 \pm 2.65 & 68.63 \pm 0.95 & 72.65 \pm 1.19 & 73.33 \pm 0.22 & - & - \\ \bottomrule
\end{tabular}
\end{adjustbox}
\end{table}

\paragraph{In-domain pre-training boosts performance, especially in low-supervised data scenarios.} 
The evaluation of linear training in Figure~\ref{fig:performance} shows that the best linear separation is given by pre-training using SimCLR on in-domain data with custom augmentations (red, dotted), exceeding ImageNet pre-training (gray, solid). These results are echoed also in the fine-tuning case, as shown in Table~\ref{tab:finetuning}. Notably for smaller supervised training datasets (fewer than 65 slides), pre-training gives a substantial boost. When significantly more supervised training data is available (100 slides or more), the gain of using pre-trained weights is diminished. This is especially clear for breast data when doing fine-tuning (Table~\ref{tab:finetuning}), where no pre-training on the full supervised dataset (216 slides) gave similar performance as using initalization from either ImageNet or SimCLR pre-trained weights. For skin, the size of the full supervised dataset (of 65 slides) is still small enough to make use of pre-trained weights a good idea.

\paragraph{Optimal dataset and view generation depend on downstream task.}
For breast data, we see in Figure~\ref{fig:performance} that the two in-domain models with different augmentations (red) gave similar performance, while for skin, different augmentations gave larger difference. Using a custom set of augmentations compared to the original SimCLR gave a significant boost in performance (dotted vs dashed). Furthermore, using Multidata as pre-training dataset that is out-of-domain pathology data (blue) gave for breast data the poorest performance, independent of augmentation, while for skin, Multidata with the original augmentations was on par or just slightly worse than the best in-domain model (blue, dashed). This corroborates the theory discussed in Section~\ref{sec:viewgen}, that the features learned during pre-training are highly dependent on what data and how view generation was performed (i.e., what augmentations were applied), and that their usefulness/relevance are dependant on the downstream task.

\paragraph{Increasing diversity of pathology data is not beneficial \textit{per se}.} 
Both tissue types were evaluated on the Multidata dataset, with two different sets of augmentations each. These augmentations where chosen either as a general approach (SimCLR original) or a dataset specific augmentation. The Multidata dataset is smaller than the others, but has much larger diversity as it contains samples from a wide range of publicly available datasets. This reduces the risk of false negatives, and could potentially create more diverse sets of features. 
Looking at the linear performance in Figure~\ref{fig:performance}, we see that the same model trained on Multidata with original augmentations (blue, dashed) performed poorly on the downstream task of breast tumor detection in sentinel lymph node tissue, but gave good results in skin tissue classification. 
\newtext{This means that task-relevant features were to a higher degree extracted for skin as compared to breast. Either the dataset lack in the features needed for successful tumor detection in lymph node tissue, or the view generation could not isolate the relevant features. We conclude that a more diverse dataset does not guarantee a generalizable model \textit{per se}, but acknowledge that diversity could potentially be beneficial if formulated correctly and used with the appropriate view generation strategy. That is,}
having a diverse dataset may increase the chance of \newtext{including} 
relevant features, but if those features are learned depends on the view generation.

\begin{figure}[!t]
    \centering
    \begin{subfigure}{.5\textwidth}
        \centering
        \includegraphics[width=0.9\linewidth]{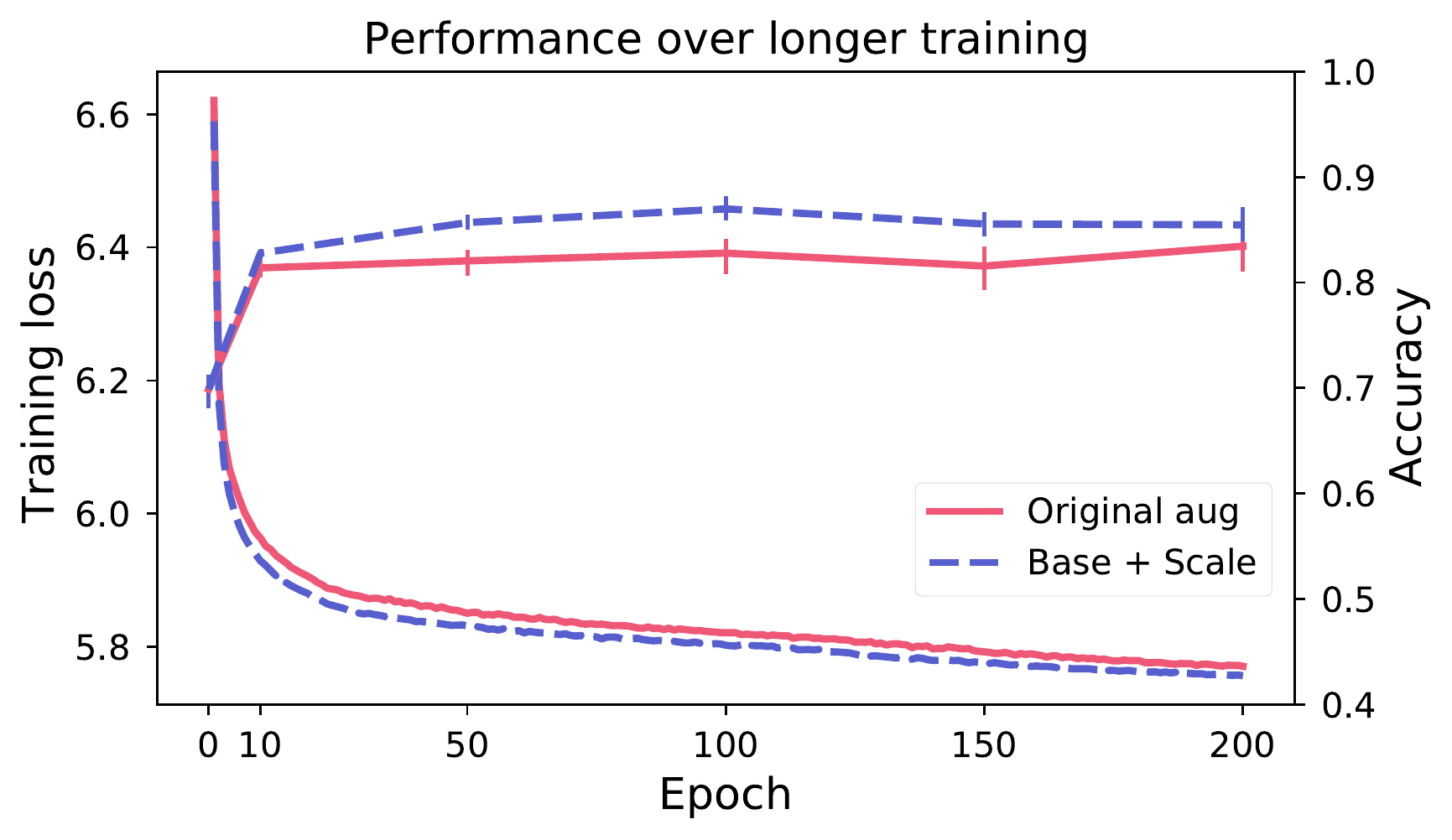}
        \caption{Breast}
    \end{subfigure}%
    \hfill
    \begin{subfigure}{.5\textwidth}
        \centering
        \includegraphics[width=0.9\linewidth]{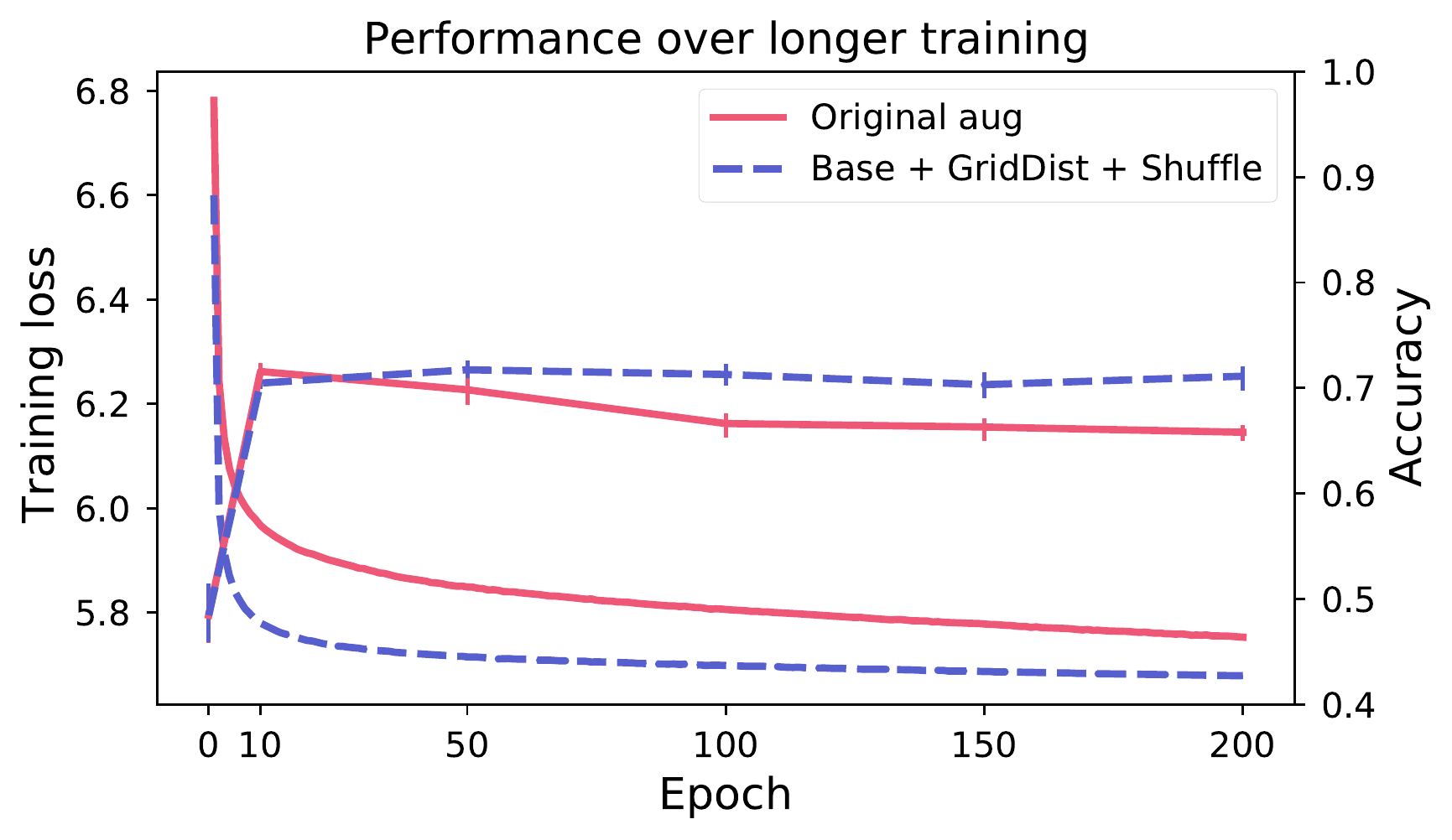}
        \caption{Skin}
    \end{subfigure}%
    \caption{Training loss and downstream accuracy evaluated for two different models on breast and skin data, respectively. Loss continues to decrease continuously during training, while the performance on the downstream application evaluated at epochs \{10, 20, 50, 100, 200\} is after the initial epochs almost constant.}
    \label{fig:longer_training}
\end{figure}

\subsection{Effects of hyper-parameters}
Large batch sizes, long training times\newtext{, and temperature scaling} have been shown to play an important role in contrastive self-supervised learning for ImageNet~\citep{chen_2020}. Here, we investigate to see if this also holds true for histology data.

\paragraph{Longer training does not improve performance.}
Figure~\ref{fig:longer_training} shows evaluation of linear performance at intervals during training, for two models respectively for breast and skin. Despite continued reduction in training loss (the model is still learning to solve the SSL objective), the performance on the downstream task is changing very little after the first 10 epochs. This indicates that the view generation fails to isolate task-relevant features, making the model rely on task-irrelevant features to solve the SSL objective (scenario shown in Figure~\ref{fig:contrastive_objective}).

\begin{figure}[t!]
    \centering
    \includegraphics[width=0.99\linewidth]{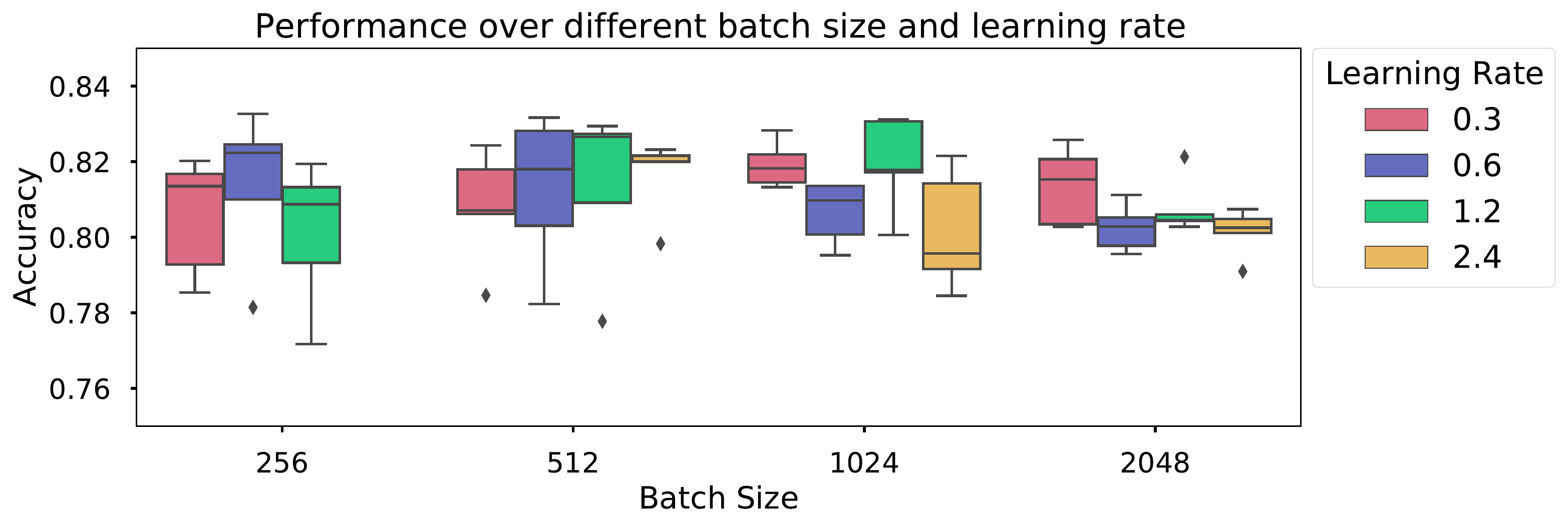}
    \caption{Performance for SimCLR trained with original augmentations for 50 epochs on breast data, using varying batch size and learning rate. Using learning rate of 2.4 with batch size 256 did not converge.}
    \label{fig:batch_size_lr}
\end{figure}

\paragraph{Large batch sizes are not needed.}
The motivation of large batch sizes is that this will form a better approximation of the true dataset distribution, wherein the separation of the positive and negatives will better reflect the true distribution. Figure~\ref{fig:batch_size_lr} shows varying batch size for a model trained on breast data, with LARS optimizer. As long as the learning rate is updated according to the size of the batch (approximately following the relationship $\textit{LearningRate} = 0.3 \cdot \textit{BatchSize}/256$ as in~\cite{chen_2020}), 
\newtext{increased batch size did not result in strictly better performance compared to smaller batch sizes. All batch sizes resulted in similar performance (at different learning rates), indicating that smaller batch sizes can be used to reach the same performance level as larger batch sizes. Similar results have been shown for volumetric medical data~\citep{chaitanya_2020}.} 

\begin{figure}[t!]
    \centering
    \includegraphics[width=0.9\linewidth]{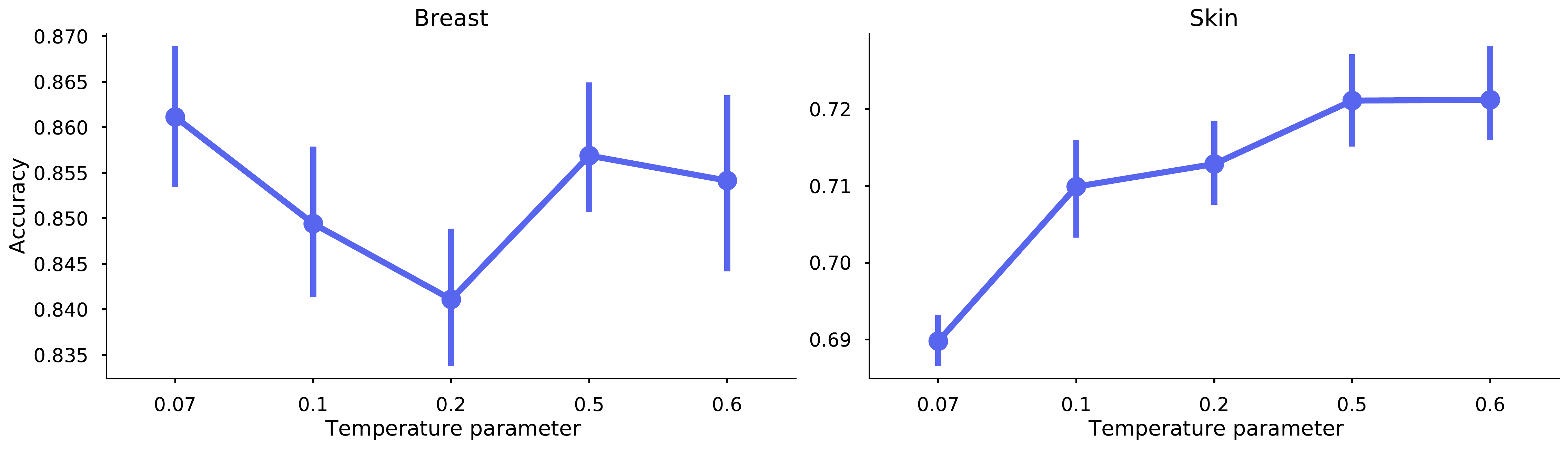}
    \caption{\newtext{Linear performance (patch-level accuracy over five trainings) for SimCLR trained with best augmentations (from Table~\ref{tab:aug}) for each datasets respectively, with varying temperature parameter. Note the different y-axes.}}
    \label{fig:temp}
\end{figure}

\paragraph{Optimal temperature scaling is dataset dependant.}
\newtext{Proper temperature scaling is important for the model to learn good representations~\citep{ciga_2022, wang_2021a}. In Figure~\ref{fig:temp}, we investigate five different values of $\tau$. The SimCLR models were trained on breast and skin data respectively, for 50 epochs and batch size 1024, using the best performing augmentations for each dataset (as of Table~\ref{tab:aug}). Similarly as the results from \cite{ciga_2022}, the different datasets may require different optimal values. However, the results show that the default value of 0.5 is a good compromise to achieve high levels of performance for both datasets.}

\section{Discussion}\label{sec:discussion}
From the results in Section~\ref{sec:exp}, we can make some interesting observations. Primarily, with correctly selected augmentations, in-domain contrastive SSL is beneficial as pre-training, especially in low-data regimes. In addition, experiments show that large batch sizes and long training times may not be needed to create pre-trained models, making model-creation more accessible. However, the results also raise concerns, which are discussed below. 

\subsection{Consequences of different dataset characteristics} 
\label{sec:data}

\begin{table}[b!]
\centering
\caption{Dataset characteristics and the consequences they have for learning with contrastive self-supervised methods.}
\label{tab:dataset_chars}
\resizebox{\textwidth}{!}{%
\begin{tabular}{@{}lccccl@{}}
\toprule
\textit{\textbf{Dataset}} &
  \multicolumn{4}{c}{\textit{\textbf{Dataset characteristics}}} &
  \textit{\textbf{Consequence}} \\ \cmidrule{2-5}
\textit{\textbf{}} &
  \multicolumn{1}{l}{\textit{\begin{tabular}[c]{@{}l@{}}Number of \\ classes\end{tabular}}} &
  \multicolumn{1}{l}{\textit{\begin{tabular}[c]{@{}l@{}}Class \\ balance\end{tabular}}} &
  \multicolumn{1}{l}{\textit{\begin{tabular}[c]{@{}l@{}}Intra-/Inter-\\ class \\ variance\end{tabular}}} &
  \multicolumn{1}{l}{\textit{\begin{tabular}[c]{@{}l@{}}Downstream-\\ target \\ isolation\end{tabular}}} &
  \textit{\textbf{}} \\ \midrule
  \vspace{2mm}
\textit{ImageNet} &
  1000 &
  \begin{tabular}[t]{@{}c@{}}Medium/ \\ Good\end{tabular} &
  Good &
  Good &
  \begin{tabular}[t]{@{}l@{}}Many classes, good balance:\\ \textbf{reduced risk of false negatives}\\ High variance and good isolation:\\ \textbf{easier view generation} \end{tabular} \\ \vspace{2mm}
\textit{Camelyon16} &
  2 &
  Low &
  Low &
  Low/Medium &
  \begin{tabular}[t]{@{}l@{}}Few classes, poor balance:\\ \textbf{higher risk of false negatives}\\ Low variance with low/\\ medium isolation:\\  \textbf{harder view generation}\end{tabular} \\
  \vspace{2mm}
\textit{AIDA-LNSK} &
  5 &
  Medium &
  Low &
  Low/Medium &
  \begin{tabular}[t]{@{}l@{}}Few classes, medium balance:\\ \textbf{slightly higher risk of false negatives}\\Low variance with low/\\ medium isolation:\\  \textbf{harder view generation}\end{tabular} \\ \bottomrule
\end{tabular}%
}
\end{table}
The results presented in Section~\ref{sec:exp} show that the heuristics derived to be optimal for ImageNet does not transfer to the other datasets. This suggests that the method is tightly coupled with the dataset. We can identify a few important dataset characteristics that affect the learning outcome, \newtext{as shown in Table~\ref{tab:dataset_chars}} 
. In the table, ImageNet is compared with the histopathology datasets of breast sentinel lymph node tissue (Camelyon16) and skin tissue (AIDA-LNSK) (presented in Section~\ref{sec:datasets}), with respect to these characteristics, and the discussion is expanded below.

\paragraph{Number of classes and class balance affect the risk of false negatives.}
When negatives are chosen as random samples from the dataset, the distribution of the classes in the dataset affects the risk of getting a high number of false negatives. With many classes and perfect balance between them, the risk of drawing negatives belonging to the same class as the anchor data point is low. In the case of ImageNet, with 1000 classes, drawing 1024 (or even up to 4096) random samples from the mini-batch as negatives, the likelihood of false negatives is low. Compare this with the histology datasets with 2 and 5 classes respectively, and where the poorer class balance further increases the risk for false negatives for the larger classes. In addition, ImageNet was collected in a supervised manner. For both histopathology data sources, the datasets used for SSL were sampled without knowledge of class labels, making the class distribution depend only on the natural occurrence of the class (in contrast to stratified sampling). 

\paragraph{Diversity in/across classes and downstream-task isolation makes view generation easier.}
As contrastive SSL aims to do instance discrimination, having data samples that are distinct makes the objective easier. Variance across classes further helps the model to learn features that separate between classes. The nature of the datasets allows ImageNet to have higher variance both within and across classes, as attributes such as viewpoints, backgrounds, lighting, occlusions, etc, can vary in a natural dataset. Even within subgroups of labels, such as dogs, we have a larger variation due to more diversity of color, texture, shape and size -- all of which, to a large extent, are constant between histopathology classes. In the ImageNet case, the augmentations applied can make use of the known variance, making the model invariant to them. The subtleties of the difference between classes in histology data makes it more challenging to find effective augmentations.

Downstream-target isolation relates to how isolated the downstream targets are on average in one image. ImageNet contains many object-centered images, containing only the class object, while both histopathology datasets have images that contain multiple classes (tissues and/or cell types). 
\newtext{Having multiple objects in an image can introduce noise. For example, }augmentations such as scale may in cases with low downstream-target isolation create a positive pair depicting two separate objects, instead of the same object in different scales. 

\subsection{How to do contrastive learning for histology?}
From what we have seen so far, the cause of the success of contrastive self-supervised learning (SSL) methods on ImageNet has been highly dependent of intrinsic properties of the data. The intricate interplay between method and data raises questions on both how to adapt the method to better accommodate the data and how to better assemble datasets that fit the method.

\paragraph{Current positive-view generation is not sufficient.}
From the results shown in Table~\ref{tab:aug} we saw that a tailored set of augmentations gives substantial improvements in downstream performance. However, we also saw in Figure~\ref{fig:longer_training} that the representations learned are, to a large extent, based on 
\newtext{features not relevant for the downstream task}. As optimal augmentations depend on both dataset and task, finding a general approach that applies to all datasets and all tasks may not be feasible. The augmentations found in this study were sub-optimal even when tailored with a specific dataset and task in mind. Creating augmentations that are strong enough to retain only label information and remove all other is not trivial, and may require extensive domain knowledge. This is further exacerbated by the fact that pathologists generally are not used to describing diagnostic criteria in terms of features suitable to formulate as image transformations. Moreover, \newtext{use of heavily tailored augmentations could be criticized as a step towards unwanted feature engineering, in the sense that expert preconceptions could constrain the self-supervised deep learning approach} 
A different direction would be to optimally learn what augmentation to apply, such as presented by \cite{tamkin_2021}. However, contrary to what \cite{tamkin_2021} suggest, the conclusion from our results is that these augmentations need to be optimized for the downstream task, not the SSL objective. A semi-self-supervised approach could therefore be an interesting future research direction. 

\paragraph{Many false-negatives gives conflicting signals}
Datasets with few classes and/or large class imbalances suffers a large risk of introducing false negatives 
\newtext{. As the mini-batch contains many samples of the same class, negatives picked from the same class are likely to occur.} This risk also increases with larger batch sizes, as the ratio between number of classes and batch size increases. This could be one explanation why there is little benefit of increased batch size for histopathology data (Figure~\ref{fig:batch_size_lr}). Having a large portion of false negatives has consequences for the learning outcome, as this prevents the model from using class-specific features to discriminate between positives and negatives (as highlighted in Figure~\ref{fig:contrastive_objective}). We can further investigate this by looking at the cosine distance between samples in a mini-batch (1024 samples) of a \newtext{SimCLR} model trained on breast data. 
A significant number of negative samples have high or very high cosine similarity ($>0.9$) with the anchor data point, an occurrence not seen for ImageNet data. Out of all negatives, $2\%$ of the anchor-negative pairs had a similarity higher than 0.9, while the same number of ImageNet is $0.0003\%$. Minimizing the number of false negatives is an important part of getting better performance~\citep{chuang_2020, chaitanya_2020}. 

\begin{newtextblock}
Methods for completely removing negatives have been presented, such as BYOL~\citep{grill_2020}, which no longer uses the contrastive objective. Some exploratory experiments in this direction using BYOL are shown in Appendix~\ref{sec:appendix_results}, Table~\ref{tab:byol_results}. 
In these experiments, BYOL does not outperform SimCLR on either of the datasets. However, further research is needed to fully understand the role of negatives for optimal self-supervised learning for histopathology.
\end{newtextblock}

\paragraph{Intrinsic properties of histology datasets may be incompatible with current methods.}
As discussed in Section~\ref{sec:data}, intrinsic properties of the datasets makes positive view generations challenging (low inter- and intra-class variance and lower target object isolation in individual images), and increases the risk of false-negatives (due to low number of classes and poor class distribution). Even if these problems could be addressed with new techniques such as better view generation and true negative sampling, questions regarding the suitability of SSL for these types of datasets remain. The SimCLR objective optimizes towards \textit{instance discrimination}. This approach is intuitive when we have a dataset where the intra-class variance consists of multiple ways of describing the same object. Being able to separate each of these instance helps give a wider distribution of the possible appearances of the object in question. In histopathology, datasets are constructed as smaller patches from whole-slide images, and where the intra-class variance consists of images showing multiple cells, where the cells actually are more or less clones of each other. A different approach of constructing these datasets may be needed, such that the downstream-target isolation becomes higher. This is indeed challenging. Taking smaller patches depicting as little as individual cells suffers even more of false negatives, and macro structures may be hard to learn. Taking larger patches/reduce resolution would include larger structures, but may include multiple tissues at once, reducing downstream-target isolation further \newtext{(visual comparison of patches sampled at different resolutions are shown in Appendix \ref{fig:example_res})}. 

\paragraph{If you understand your data, then contrastive self-supervised learning can still be useful.} 
Despite the above mentioned limitations for contrastive SSL used on histology data, the current setup may still bring value for specific histology applications. For example, the reduced need for large batch sizes and long training times makes in-domain pre-training using contrastive SSL accessible to a larger community. Furthermore, by keeping the dataset characteristics from Table~\ref{tab:dataset_chars} in mind, risk factors for the dataset in question may be early identified. By understanding the inter- and intra-variance of the (latent) classes of the dataset, augmentations may be formulated that are better tailored to the specific downstream application in mind, compared to naively using those optimized for ImageNet. The risk of false negatives might be possible to mitigate during the data collection, for example by controlling \newtext{the field of view by changing} the resolution in which the patches are sampled. In combination with the results shown in Table~\ref{tab:finetuning} that shows that in-domain SimCLR pre-training does boost performance, contrastive self-supervised learning can indeed be a way to reduce the need for labeled data for histopathology applications.

\subsection{Limitations and Future work}

This paper aims to evaluate if and how contrastive self-supervised methods can be used to reduce the needed amount of labeled data for the target histopathology application. The scope of the study was limited to three datasets and two classification tasks, and where SimCLR was chosen as representative method among all contrastive self-supervised methods. Restricted by the challenges and limitations of pre-training models for \textit{one} downstream task, we did not evaluate the generalization of the SSL models by evaluating one pre-trained model on multiple downstream tasks, with one exception (SimCLR Multidata with original augmentations was used as pre-trained model for both downstream tasks). Despite the restricted scope, we believe that the results may give guidance when applied to an extended domain.

The results from this study show that contrastive self-supervised methods have the potential, if applied correctly, to reduce the need for labeled target data. However, they also show that the method is still sub-optimal with respect to the specific data characteristics of histopathology. There is therefore room for improvement, but the challenges of creating informative positives and reduce false negatives are not trivial to solve. Creating informative positives may be easier with deepened understanding of what features should be considered task-relevant for a given downstream task. The problem of false negatives could potentially be solved by selecting negatives in a non-random way, potentially taking a semi-supervised approach. We hope that this work will inspire interesting future research that take a holistic approach, considering the interplay between dataset and method.

\section{Conclusions}

In this paper, we have evaluated contrastive self-supervised learning on histopathology data. Effective contrastive self-supervised learning with respect to a particular downstream task requires two criteria to be fulfilled, namely that the shared information between the positive views is high, and that the false negative rate is low. Our study shows that both these criteria are challenging to fulfill for histopathology applications, due to the characteristics of the datasets. Furthermore, we have shown that the explicit and implicit heuristics used for ImageNet does not necessarily apply in the domain of histopathology. We conclude that SSL for histopathology cannot be considered and used under the same assumptions as for natural images, and that in-depth understanding of the data is essential for training self-supervised models for histopathology applications.


\acks{This work was supported by the Wallenberg AI and Autonomous Systems and Software Program (WASP-AI), the research environment ELLIIT, AIDA Vinnova grant 2017-02447, and Linköping University Center for Industrial Information Technology (CENIIT). The computations were enabled by the supercomputing resource Berzelius provided by National Supercomputer Centre at Linköping University and the Knut and Alice Wallenberg foundation. We also like to thank our colleague Jesper Molin (Sectra) for comments on the manuscript.}

%
\ethics{The work follows appropriate ethical standards in conducting research and writing the manuscript, following all applicable laws and regulations regarding treatment of animals or human subjects.}

\coi{We declare that we do not have any conflicts of interest.}

\bibliography{main}

\newpage
\appendix 

\section{Datasets}
\label{sec:appendix_data}

In Table~\ref{tab:dataset_details} the number of slides and patches are given for the different datasets. All patches were extracted at 0.5 microns per pixel resolution with size 256x256 pixels. The unsupervised dataset were sampled without overlap in a uniform grid. For breast, maximally 1000 samples per slide was randomly selected from slides in the training set, resulting in approximately 270k patches. Similarly for skin, the unsupervised training set consists of samples selected without using labels, resulting in approximately 270k patches. Potential sampling points were found by, in tissue regions, extracting samples from a random, uniform grid. From these candidates, 1000 samples per slide were extracted at random. 

The supervised training, validation and test set for breast follows the patch coordinates of PatchCamelyon, but was resampled at the above mentioned resolution and size. For skin, the supervised datasets were constructed as follows. All slides were sampled in a uniform grid with 50\% overlap, resulting in total of 1.3 million candidate patches. From these candidates, a subset where selected for each dataset, with the follow criteria.
For the training set, the large class imbalance was mitigated by selecting patches such that for each class, the total number of patches was \texttt{min}(75 000, \texttt{all available}). This resulted in approx. 320k patches. For validation, 700 patches from each class were randomly selected from the slides in the validation set. Similarly, 3700 patches from each class were randomly selected from the slides in the test set to formulate the test patches. The validation and test sets are therefore class balanced. There is no patient overlap between the supervised datasets.

\begin{table}[ht]
\centering
\caption{Number of whole-slide images (WSIs) and patches for breast and skin data.}
\label{tab:dataset_details}
\begin{adjustbox}{width=\textwidth}
\begin{tabular}{llccc}
\hline  \multicolumn{2}{l}{\textit{Unsupervised}} & \textit{Training} & \textit{Validation} & \textit{Test} \\ \cmidrule(lr){1-2}\cmidrule(lr){3-3}\cmidrule(lr){4-4}\cmidrule(lr){5-5}
\textit{\textbf{Breast}} & \textit{\# WSI}     & 270                  & N/A                  & N/A                  \\
                         & \textit{\# patches} & 265048               & N/A                  & N/A                  \\
                         &                     & \multicolumn{1}{l}{} & \multicolumn{1}{l}{} & \multicolumn{1}{l}{} \\
\textit{\textbf{Skin}}   & \textit{\# WSI}     & 65                   & N/A                  & N/A                  \\
                         & \textit{\# patches} & 271675               & N/A                  & N/A                  \\
\multicolumn{2}{l}{\textit{Supervised}} & & & \\ \cmidrule(lr){1-2}\cmidrule(lr){3-3}\cmidrule(lr){4-4}\cmidrule(lr){5-5}
\textit{\textbf{Breast}} &
  \textit{\# WSI} &
  216 / 100 / 50 / 20 / 10 &
  54 &
  129 \\
\textit{} &
  \textit{\# patches} &
  262144 / $\sim$120000 / $\sim$60000 / $\sim$25000 / $\sim$10000 &
  32768 &
  32768 \\
                         &                     & \multicolumn{1}{l}{} & \multicolumn{1}{l}{} & \multicolumn{1}{l}{} \\
\textit{\textbf{Skin}}   & \textit{\# WSI}     & 65 / 50 / 20 / 10    & 8                    & 23                   \\
\textit{} &
  \textit{\# patches} &
  317243 / $\sim$235000 / $\sim$95000 / $\sim$50000 &
  3500 &
  18500 \\ \hline
                         &                     & \multicolumn{1}{l}{} & \multicolumn{1}{l}{} & \multicolumn{1}{l}{}
\end{tabular}
\end{adjustbox}
\end{table}

\section{Augmentations}
\label{sec:appendix_aug}
\begin{figure}[H]
    \centering
    \includegraphics[width=\linewidth]{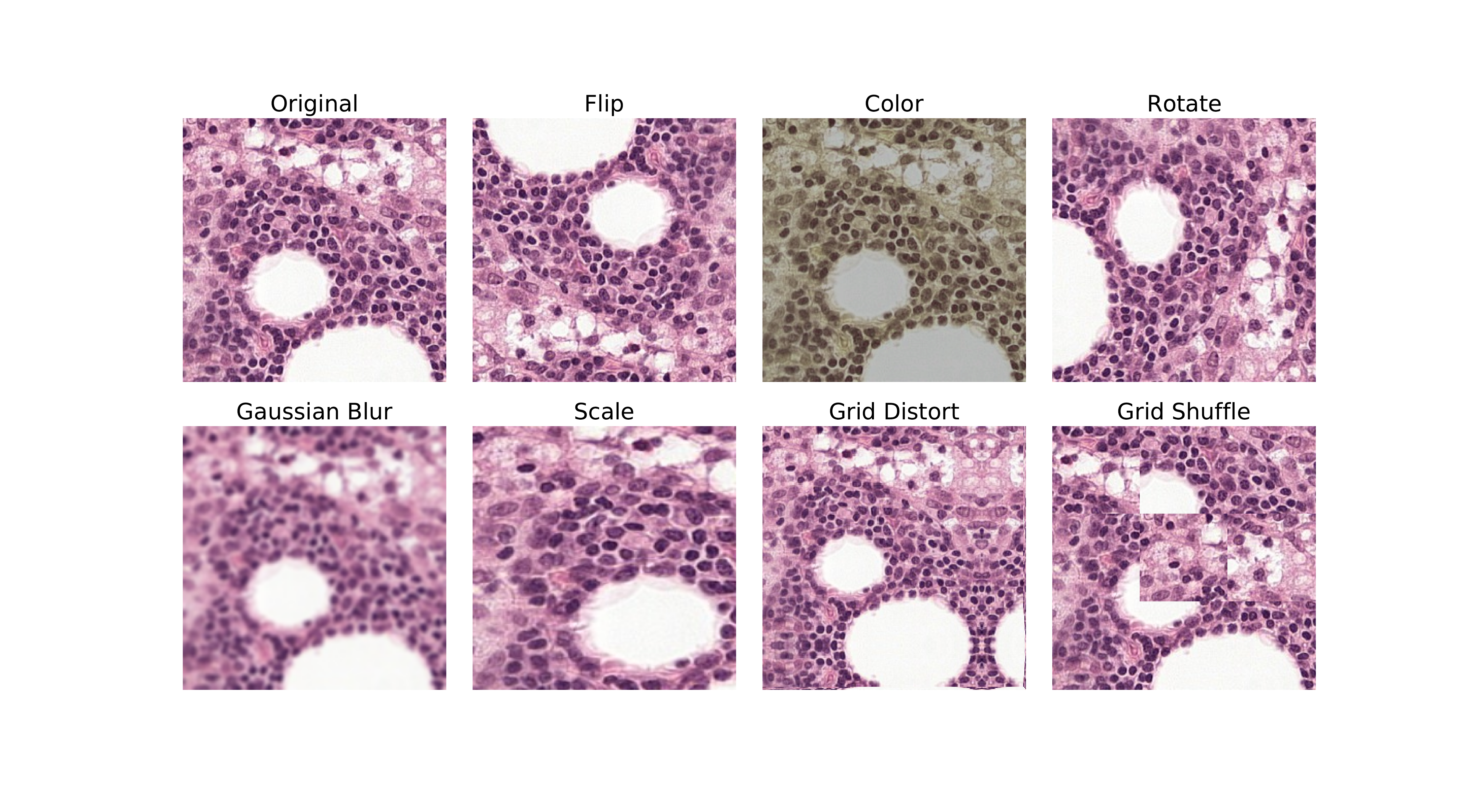}
    \caption{Examples of augmentations}
    \label{fig:ex_aug}
\end{figure}

The augmentations applied were done using Pytorch Transforms (\url{https://pytorch.org/vision/stable/transforms.html}) or with Albumentations~\citep{buslaev_2020}. The implementation of Gaussian blur was taken from: \url{https://github.com/facebookresearch/moco}. Examples are shown in Figure~\ref{fig:ex_aug}, and implementation details in Table~\ref{tab:augmentations}.

\begin{table}[H]
\centering
\caption{List of applied augmentations, with corresponding parameters and probability of application.  }
\label{tab:augmentations}
\begin{tabular}{@{}llcc@{}}
\toprule
 &  Transformation &  Params &  Probability \\ \midrule
\ldelim\{{7}{*}[\parbox{30mm}{\textit{Base}}] & \textit{Random Crop} &  size: 224x224 &  1.0 \\
 &  \textit{Flip} &  - &  0.5 \\
 &  \textit{Rotation (fixed 90 degrees)} &  - &  0.5 \\
 &  \multirow[t]{4}{*}{\textit{Color Jitter}} &  brightness: 0.8 & \multirow[t]{4}{*}{0.8} \\
 & & contrast: 0.8 & \\ 
 & & saturation: 0.8 &\\ 
 & & hue: 0.2 &  \\ 
\ldelim\{{2}{*}[\parbox{30mm}{\textit{SimCLR original}}] &
  \textit{Scale} &
  scale: \{0.2, 0.95\} -- 1.0 &
  1.0 \\
 &
  \textit{Gaussian Blur} &
  sigma: 0.1--2.0 &
  \{0, 0.5\} \\ 
 &  \multirow[t]{3}{*}{Grid Distortion} &  num\_steps: 9 &  \multirow[t]{3}{*}{\{0, 0.5\}} \\
 & & distort\_limit: 0.2 & \\ 
 & & border\_mode: 2 &  \\
 &  \textit{(Grid) Shuffle} &  grid: (3,3) &  \{0, 0.5\} \\ \bottomrule
\end{tabular}
\end{table}

\section{Results}
\label{sec:appendix_results}

Tables~\ref{tab:linear_breast} and \ref{tab:linear_skin} show the results presented in Figure~\ref{fig:performance} in numerical form.

\begin{table}[H]
\centering
\tiny
\caption{Linear performance for patch-level PatchCamelyon, shown as accuracy (\%) and patch-level AUC}
\label{tab:linear_breast}
\begin{tabular}{@{}llccccc@{}}
\toprule
 & \textbf{Model}                       & \multicolumn{5}{c}{\textbf{Supervised size (\#slides)}}            \\ \cmidrule{3-7}
 &                                      & 10          & 20          & 50          & 100         & 216         \\ \midrule
\textit{Accuracy (\%)} 
  & \textit{Supervised}   & $73.3 \pm 4.8$ & $77.8 \pm 2.3$ & $82.5 \pm 0.9$ & $89.3 \pm 1.7$ & $93.1 \pm 0.2$ \\
 & \textit{ImageNet Supervised} &  $74.9 \pm 4.9$ &  $80.5 \pm 3.1$ &  $84.2 \pm 0.5$ &  $85.6 \pm 0.4$ &  $86.1 \pm 0.1$ \\
 & \textit{SimCLR Breast, Original}     & $76.7 \pm 3.9$ & $77.8 \pm 2.7$ & $83.4 \pm 2.6$ & $86.5 \pm 0.5$ & $88.7 \pm 0.3$ \\
 & \textit{SimCLR Breast, Base + Scale} & $78.0 \pm 5.0$ & $80.2 \pm 1.8$ & $85.5 \pm 1.8$ & $87.4 \pm 0.8$ & $89.6 \pm 0.1$ \\
 & \textit{Multidata, Original}         & $72.6 \pm 2.9$ & $73.6 \pm 2.7$ & $79.6 \pm 2.4$ & $82.3 \pm 1.4$ & $84.3 \pm 0.1$ \\
 & \textit{Multidata, Base + Scale}     & $73.2 \pm 3.2$ & $73.4 \pm 3.2$ & $79.5 \pm 1.6$ & $81.8 \pm 1.6$ & $83.1 \pm 0.1$ \\ \midrule
\multicolumn{1}{c}{\textit{AUC}} 
& \textit{Supervised}         & $0.805 \pm 0.055$ & $0.855 \pm 0.027$ & $0.900 \pm 0.010$ & $0.960 \pm 0.009$ & $0.979 \pm 0.001$ \\
  & \textit{ImageNet Supervised} & $0.838 \pm 0.046$ & $0.887 \pm 0.024$ & $0.918 \pm 0.004$ & $0.928 \pm 0.002$ & $0.935 \pm 0.000$ \\
 & \textit{SimCLR Breast, Original}     & $0.830 \pm 0.072$ & $0.865 \pm 0.057$ & $0.920 \pm 0.032$ & $0.947 \pm 0.008$ & $0.957 \pm 0.001$ \\
 & \textit{SimCLR Breast, Base + Scale} & $0.842 \pm 0.074$ & $0.884 \pm 0.031$ & $0.929 \pm 0.019$ & $0.944 \pm 0.010$ & $0.957 \pm 0.000$ \\
 & \textit{Multidata, Original}         & $0.820 \pm 0.041$ & $0.837 \pm 0.044$ & $0.888 \pm 0.021$ & $0.910 \pm 0.015$ & $0.921 \pm 0.001$ \\
 & \textit{Multidata, Base + Scale}     & $0.827 \pm 0.047$ & $0.839 \pm 0.053$ & $0.892 \pm 0.015$ & $0.909 \pm 0.012$ & $0.909 \pm 0.001$ \\ \bottomrule
\end{tabular}
\end{table}

\begin{table}[H]
\centering
\caption{Linear performance (patch-level accuracy (\%)), AIDA-LNSK}
\label{tab:linear_skin}
\begin{adjustbox}{width=\textwidth}
\begin{tabular}{@{}llcccc@{}}
\toprule
 & \textbf{Model}                       & \multicolumn{4}{c}{\textbf{Supervised size (\#slides)}}            \\ \cmidrule{3-6}
 &                                      & 10          & 20          & 50          & 65         \\ \midrule
\textit{Accuracy (\%)} 
  & \textit{Supervised}                 & $61.4 \pm 3.9$ & $66.6 \pm 1.8$ & $69.6 \pm 0.7$ & $71.4 \pm 0.3$ \\
 & \textit{ImageNet Supervised}         & $64.2 \pm 3.9$ & $66.6 \pm 1.2$ & $68.3 \pm 0.4$ & $69.0 \pm 0.2$ \\
 & \textit{SimCLR Skin, Original}       & $59.9 \pm 1.8$ & $62.2 \pm 1.4$ & $65.8 \pm 0.8$ & $66.7 \pm 0.6$ \\
 & \textit{SimCLR Skin, Base + GridDist + Shuffle} & $65.2 \pm 3.0$ & $67.3 \pm 1.7$ & $71.1 \pm 1.3$ & $72.0 \pm 0.4$  \\
 & \textit{Multidata, Original}         & $63.9 \pm 5.1$ & $67.4 \pm 1.7$ & $70.3 \pm 0.7$ & $71.2 \pm 0.2$ \\
 & \textit{Multidata, Base + GridDist + Shuffle} & $62.7 \pm 4.9$ & $65.9 \pm 1.5$ & $69.0 \pm 0.5$ & $69.5 \pm 0.3$ \\ \bottomrule
\end{tabular}
\end{adjustbox}
\end{table}

Table~\ref{tab:batch_size} shows the results presented in Figure~\ref{fig:batch_size_lr} in numerical form.

\begin{table}[h]
\centering
\caption{\newtext{Linear performance (patch-level accuracy (\%)) of SimCLR model trained with original augmentations on breast data for 50 epochs, evaluated on PatchCamelyon (50 slides subset).}}
\label{tab:batch_size}
\begin{tabular}{ccccc}
\toprule
\multicolumn{1}{r}{\textbf{Batch Size}} & \multicolumn{4}{c}{\textbf{Learning Rate}} \\
\multicolumn{1}{r}{} & 0.3 & 0.6 & 1.2 & 2.4 \\ \midrule
\textit{256} & $0.81 \pm 0.02$ & $0.81 \pm 0.02$ & $0.80 \pm 0.02$ & - \\
\textit{512} & $0.81 \pm 0.02$ & $0.81 \pm 0.02$ & $0.81 \pm 0.02$ & $0.82 \pm 0.01$ \\
\textit{1024} & $0.82 \pm 0.01$ & $0.81 \pm 0.01$ & $0.82 \pm 0.01$ & $0.80 \pm 0.02$ \\
\textit{2048} & $0.81 \pm 0.01$ & $0.80 \pm 0.01$ & $0.81 \pm 0.01$ & $0.80 \pm 0.01$ \\ \bottomrule
\end{tabular}
\end{table}

\begin{newtextblock}
In Table~\ref{tab:byol_results}, results are shown comparing SimCLR model to BYOL. The models were trained on in-domain data, with either the original SimCLR augmentations or the best performing augmentations from Table~\ref{tab:aug}. The models were trained for 200 epochs, with batch size 1024. The SimCLR model was trained with learning rate 1.2, but BYOL was found to be needed lower learning rate 0.2. The results are presented as the mean linear patch-wise accuracy of the supervised subset of 50 slides, evaluted over 5 runs. 
\end{newtextblock}

\begin{table}[h]
\centering
\caption{\newtext{Comparison between SimCLR method and BYOL, trained for 200 epochs, evaluated on linear patch-level accuracy of the supervised subset of 50 slides.}}
\label{tab:byol_results}
\begin{tabular}{@{}llcc@{}}
\toprule
\textbf{Dataset} & \textbf{Method} & \multicolumn{2}{c}{\textbf{Augmentation}} \\
\textbf{} & \textbf{} & Original & Best. \\ \midrule
\textit{Breast} & \textit{SimCLR} & $0.83 \pm 0.03$ & $0.85 \pm 0.02$ \\
\textit{} & \textit{BYOL} & $0.79 \pm 0.01$ & $0.83 \pm 0.00$ \\
\textit{Skin} & \textit{SimCLR} & $0.66 \pm 0.01$ & $0.71 \pm 0.01$ \\
\textit{} & \textit{BYOL} & $0.48 \pm 0.03$ & $0.48 \pm 0.03$ \\ \bottomrule
\end{tabular}
\end{table}

\begin{figure}[h]
    \centering
    \begin{subfigure}[b]{0.45\textwidth}
        \centering
        \includegraphics[width=\textwidth]{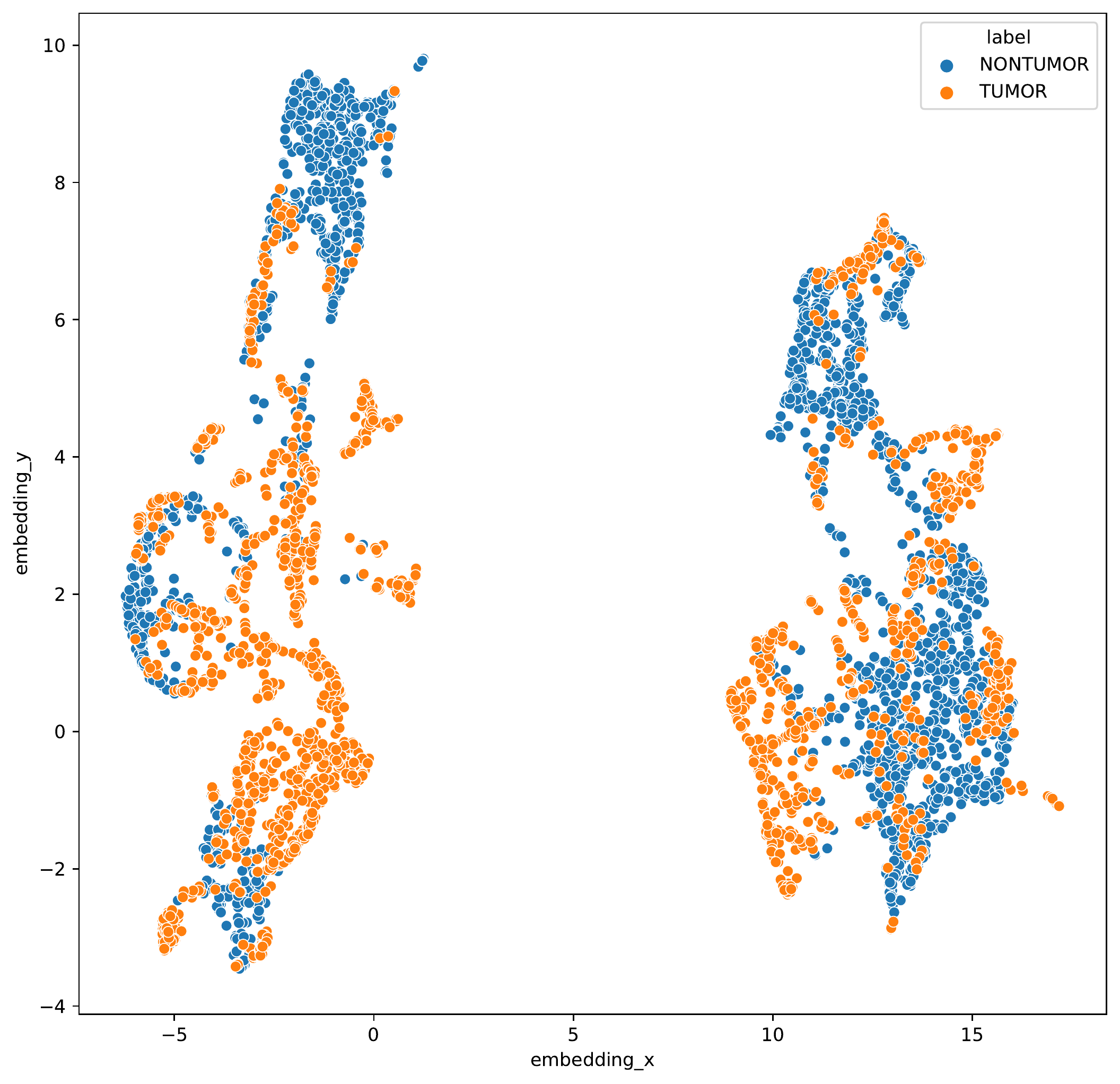}
        \caption{SimCLR trained with Original augmentations on breast data.}
    \label{fig:umap1}
    \end{subfigure}
    \hfill
    \begin{subfigure}[b]{0.45\textwidth}
        \centering
        \includegraphics[width=\textwidth]{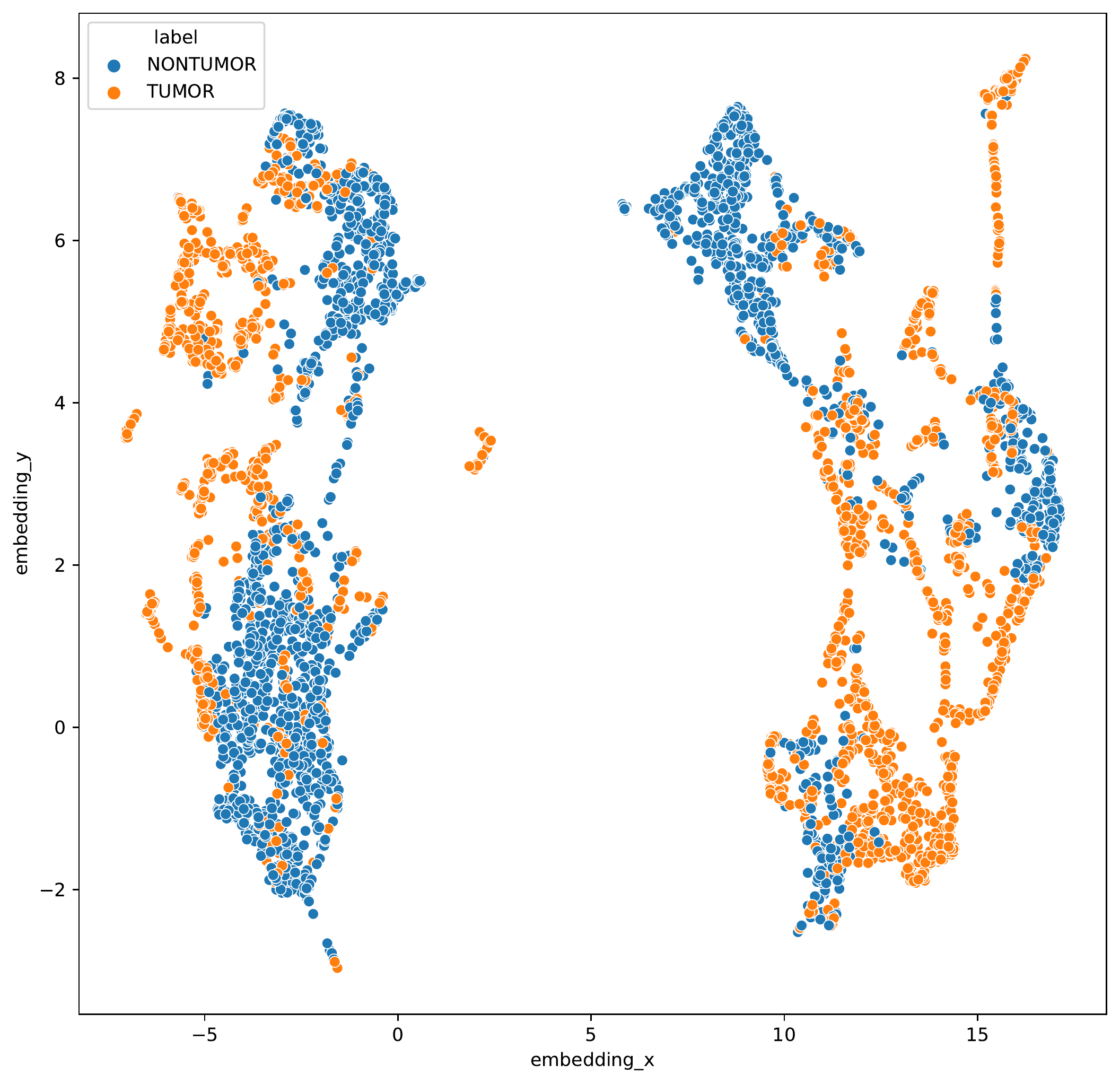}
        \caption{SimCLR trained with Base + Scale augmentations on breast data.}
        \label{fig:umap2}
    \end{subfigure}
    \vskip\baselineskip
    \begin{subfigure}[b]{0.45\textwidth}
        \centering
        \includegraphics[width=\textwidth]{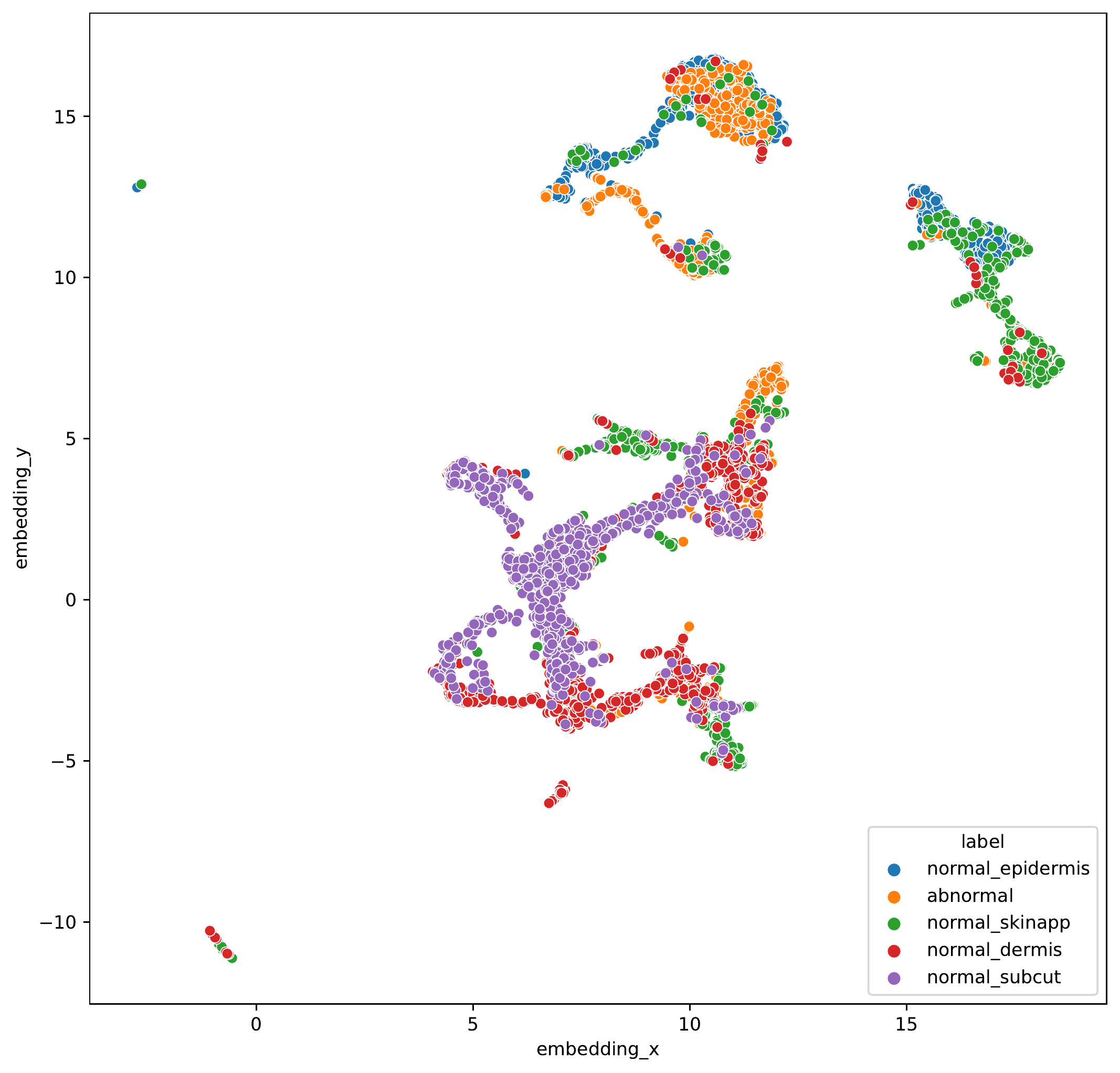}
        \caption{SimCLR trained with Original augmentations on skin data.}
        \label{fig:umap3}
    \end{subfigure}
    \hfill
    \begin{subfigure}[b]{0.454\textwidth}   
        \centering 
        \includegraphics[width=\textwidth]{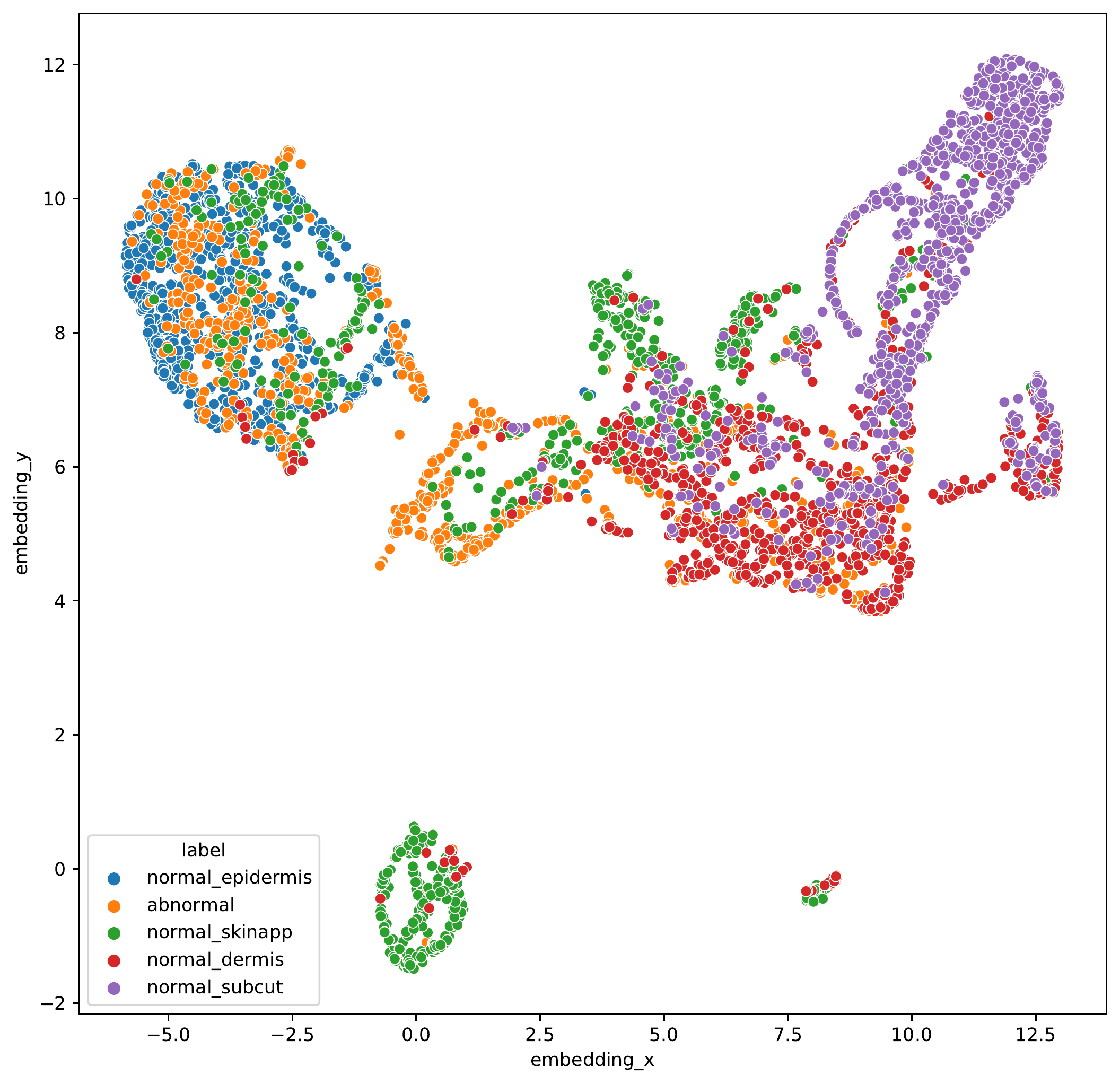}
        \caption{SimCLR trained with Base + Grid shuffle + Distort on skin data.}
        \label{fig:umap4}
    \end{subfigure}
    
    \caption{\newtext{UMAP~\citep{mcinnes_2018} visualizations of encoder embeddings of models trained on in-domain data for 200 epochs, evaluated on 5000 randomly selected samples from the test sets (class balanced). Top row: breast data, bottom row: skin data.}}
    \label{fig:umaps}
\end{figure}

\section{Discussion}
\label{sec:appendix_discussion}
\begin{figure}[h]
    \centering
    \begin{subfigure}[b]{0.9\textwidth}
        \centering
        \includegraphics[width=\textwidth]{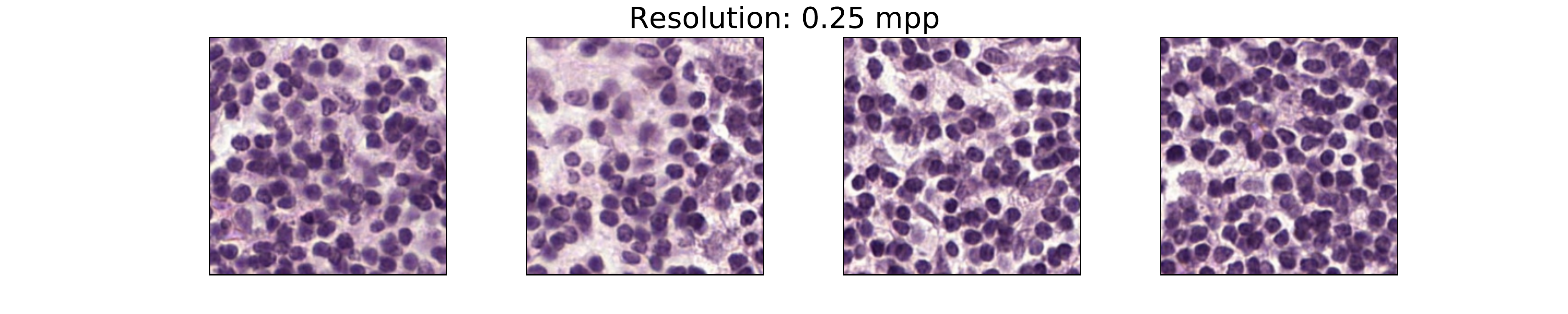}
    \end{subfigure}
    \hfill
    \begin{subfigure}[b]{0.9\textwidth}   
        \centering 
        \includegraphics[width=\textwidth]{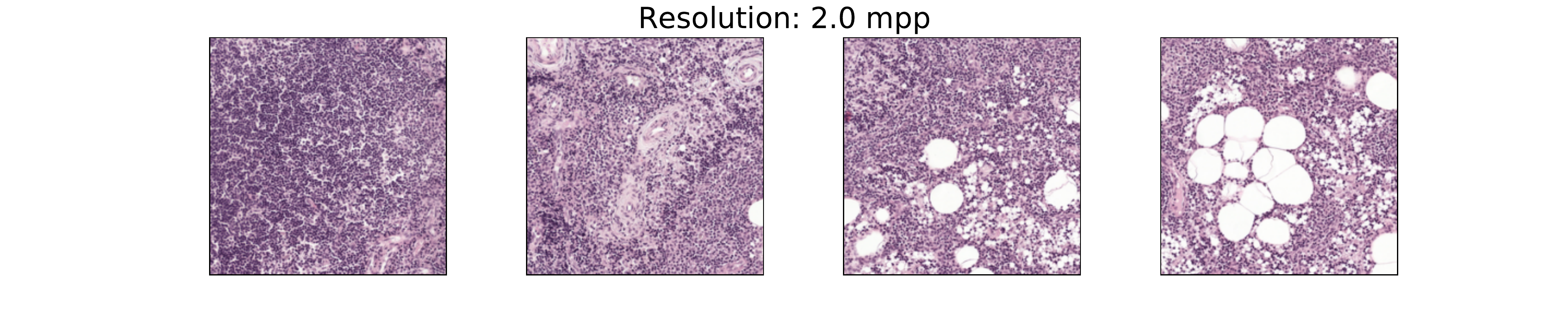}
    \end{subfigure}
    \caption{\newtext{Examples neighboring of patches extracted at different resolutions, with constant patch size. The different fields of view and level of details visible impacts what features the contrastive model will learn. }}
    \label{fig:example_res}
\end{figure}

\end{document}